\documentclass[
 aip,
jcp,
amsmath,amssymb,
preprint,%
]{revtex4-1}

\usepackage{graphicx}
\usepackage{comment}
\usepackage{dcolumn}
\usepackage{bm}
\usepackage{epsf}
\usepackage{color}

\begin{document}
\input{epsf}
\preprint{APS/123-QED}

\title{Low temperature HD + $ortho$-/$para$-H$_2$ 
inelastic scattering of astrophysical interest}

\author{Renat A. Sultanov$^{1,2}$\footnote{Electronic mail: rasultanov@stcloudstate.edu;\ r.sultanov2@yahoo.com}}
\affiliation{$^{1)}$Instituto de F\'isica Te\'orica, UNESP $-$ Universidade Estadual Paulista, 01140 S\~ao Paulo, SP, Brazil\\
$^{2)}$Department of Information Systems, BCRL \& Integrated Science and Engineering Laboratory
Facility \it{(ISELF)} at St. Cloud State University,  St. Cloud, MN, USA}

\author{Dennis Guster$^{2}$\footnote{Electronic mail: dcguster@stcloudstate.edu}}

\author{S. K. Adhikari$^{1}$\footnote{Electronic mail: adhikari@ift.unesp.br;\ http://www.ift.unesp.br/users/adhikari}}

\date{\today}
\begin{abstract} State-selected total cross sections and thermal rate 
coefficients are computed for the HD + $ortho$-/$para$-H$_2$ rotational 
energy transfer collision at low temperatures: 
2 K $\lesssim$ T $\lesssim$ 300 K. A modified H$_2$-H$_2$ 
potential energy surface (PES) devised by Hinde is used for this pure 
quantum-mechanical dynamical computation. A comparison of the new 
results for the HD + $ortho$-/$para$-H$_2$ scattering problem and previous 
calculations computed with the use of other older PESs is presented and 
discussed. \end{abstract}  \maketitle

\section{Introduction} \label{sec:intro}
Elastic and inelastic collisions between atoms and molecules and/or between
molecules and molecules are of great scientific interest in the fields of
physical-chemistry and chemical-physics. The reason is that such processes can provide valuable
information about fundamental interactions between different chemical species, their
chemical properties, their energy transfer quantum dynamics and many other properties.
The pioneering studies focussed on 
light atoms and molecules, because of their simple nature.  
For three- and four-atomic systems with a small number of electrons,
the potential energy surface  (PES) can be computed with relatively high accuracy
\cite{boothroyd,karl,schwenke88,hinde,ks,belof}. 
Consequently, for these systems one can then more easily test different dynamical methods, such
as, classical, semi-classical, quasi-classical trajectory,
and pure quantum-mechanical computational formulations and compare the 
results with available experimental data in a controlled fashion.
The test methods devised could then
be applied  to more complex many-atomic systems, wherein a controlled comparison is not possible.
Among these small systems the  four-atomic H$_2$+H$_2$ and H$_2$+HD  
scattering processes have attracted
significant attention not only in chemical physics but also in astrophysics.
In astrophysics H$_2$ and HD  play an important role, because  of their 
abundance   in the  molecular cloud of the universe \cite{book1,cmb}. Together 
with the H$_2$ + H$_2$ collision, the HD + H$_2$ collision
is also of significant importance in the astrophysics of the early 
universe. Specifically it is important
in  the modeling of pre-galactic clouds and planetary 
atmospheres; in  the cooling of 
primordial gas and in the formation of   
stars \cite{bryan08,varshal,schaefer,flower1,flower99,chu,dalgarno}.

In Ref. \cite{zarur74} a semiclassical treatment of H$_2$-H$_2$ scattering is formulated.  
In Ref. \cite{green75} the author developed and 
applied a rigid rotor model to study rotational excitation in H$_2$-H$_2$
by applying quantum
close-coupling scattering calculations. In this approach
the distance between the hydrogen atoms in both H$_2$ molecules was fixed
at a constant value based on avarage. This model was applied to many 
different atomic and molecular systems, see, for example \cite{flower99,my1,my2,my22}.
The main goal of the work \cite{green75} was to compute rotational thermal rate coefficients
in the H$_2$+H$_2$ system at low temperatures of astrophysical interest.
Quantum-mechanical close-coupling calculations for three-dimensional
collisions of $para$-H$_2$ and $ortho$-H$_2$ with HD are performed in 
Ref. \cite{schaefer,flower99,my2,my22,aip2012}, where the HD-H$_2$ potential 
is derived from the H$_2$-H$_2$ potential. A  quantum dynamical study of   
H$_2$-H$_2$ collisions is reported for both $ortho$- and $para$-H$_2$ in Refs. 
\cite{clary02,guo03,otto5}. In Ref.  \cite{montero14}, the authors considered H$_2$+H$_2$ and 
D$_2$+D$_2$ rotational inelastic scatterings with the use of the H$_2$-H$_2$ potential energy
surface (PES) from Ref. \cite{karl}.
In Ref.  \cite{clary02,bala} a full six-dimensional scattering computation has been performed
taking into account vibrational relaxation in the H$_2$+H$_2$ collision. In that study the H$_2$-H$_2$ 
PES from Ref.  \cite{boothroyd} was used and its anisotropy properties
have been studied at low temperatures: 20 K $\lesssim$ T $\lesssim$ 300 K.
HD+HD scattering has been studied experimentally in Ref. \cite{hdhd77} and theoretically
in \cite{my_hdhd} for a wide range of collision energies.
A comprehensive computational and experimental study of total cross section 
in H$_2$-H$_2$, D$_2$-D$_2$, and HD-HD 
scattering for both ortho and para H$_2$ and D$_2$ has been reported in Ref. \cite{johnson79}.
Measurements of energy transfer rates in HD+HD \cite{chandler86} and H$_2$+H$_2$
\cite{chandler88,montero05} collisions have also been performed.

However, realistic theoretical investigations  of the low-energy HD+H$_2$ collision are 
lacking, although preliminary  quantum calculation of this process has been reported 
in Refs. \cite{schaefer,flower99}. Schaefer \cite{schaefer} calculated rate coefficients
for the excitation of HD by  H$_2$, 
for the low-lying rotational levels using a modified older potential for HD and H$_2$.
Flower \cite{flower99} performed an improved calculation  of HD-H$_2$ scattering with 
Schwenke's H$_2$-H$_2$ PES using a larger rotational basis set \cite{schwenke88}.
In this paper we report an improved calculation of this problem using a realistic 
HD-H$_2$ PES derived from Hinde's recent H$_2$-H$_2$ PES \cite{hinde}.
In two papers \cite{my2,my22} the PES from work \cite{boothroyd} has been applied
together with a pure quantum-mechanical dynamical approach.
The surprising thing is, that the results of works \cite{my2,my22} are closer
to the results of older work \cite{schaefer} than to \cite{flower99}. Therefore, there is
a need to carry out new computations with newer PES between HD and H$_2$.

While the exchange symmetry is broken in HD+H$_2$ it still posses many 
similarities with the H$_2$+H$_2$ system. The PESs of H$_2$-H$_2$ and HD-H$_2$ 
should  basically be the same six-dimensional function. This fact 
follows from the general Born-Oppenheimer approach \cite{BO}. 
At the same time the two collisions: H$_2$+H$_2$ and H$_2$+HD, 
should have rather different scattering outputs. This is because the 
H$_2$ and HD molecules have fairly different rotational 
constants, internal symmetries and as 
a result a different rotational-vibrational spectrum. 
The HD-H$_2$ PES can be derived from the H$_2$-H$_2$  PES by 
shifting the center of mass (c.m.) of the H$_2$ molecule to the  
c.m. of the HD molecule. 
Once the exchange symmetry is 
broken in H$_2$-H$_2$ by replacing the H with the D atom in one of the H$_2$'s then
one has the new HD-H$_2$ PES. In this fashion, we constructed the 
HD-H$_2$ PES from the H$_2$-H$_2$ PES of Hinde \cite{hinde}
employing all parts of the full 
HD-H$_2$ interaction including the HD's dipole moment.
Using this HD-H$_2$ PES we carried out pure quantum-mechanical
calculations for inelastic collisions of rotationally excited HD 
and H$_2$ molecules, i.e. the process:
\begin{equation}
\mbox{HD}(j_1)+\mbox{H}_2(j_2)\rightarrow \mbox{HD}(j'_1)+\mbox{H}_2(j'_2).
\label{eq:h2hd}
\end{equation}
The scattering cross sections and their corresponding thermal rate 
coefficients are computed using a non-reactive quantum-mechanical 
close-coupling approach. The four-atomic system is 
shown in Fig.~\ref{fig:fig1}.

In Sec. II we briefly describe the quantum-mechanical approach used in this paper.
Sec. III includes the computational results. We 
compare the  cross-sections and rates with those of other authors 
\cite{schaefer,flower99}, and our previous  
calculations \cite{my2,my22}, where a different HD-H$_2$ PES,  
derived from the well-known Boothroyd-Martin-Keogh-Peterson (BMKP) H$_2$-H$_2$ PES
\cite{boothroyd}, was used. Discussion and conclusions are provided in Sec. IV.
The corresponding procedure to obtain a modified HD-H$_2$ PES from the existing 
H$_2$-H$_2$ surface \cite{hinde} is presented in Sec. \ref{sec:HD_PES}.
Atomic units $(e=m_{e}=\hbar=1)$ are used throughout this paper.  

\section{Quantum-mechanical approach}
\label{sec:method}
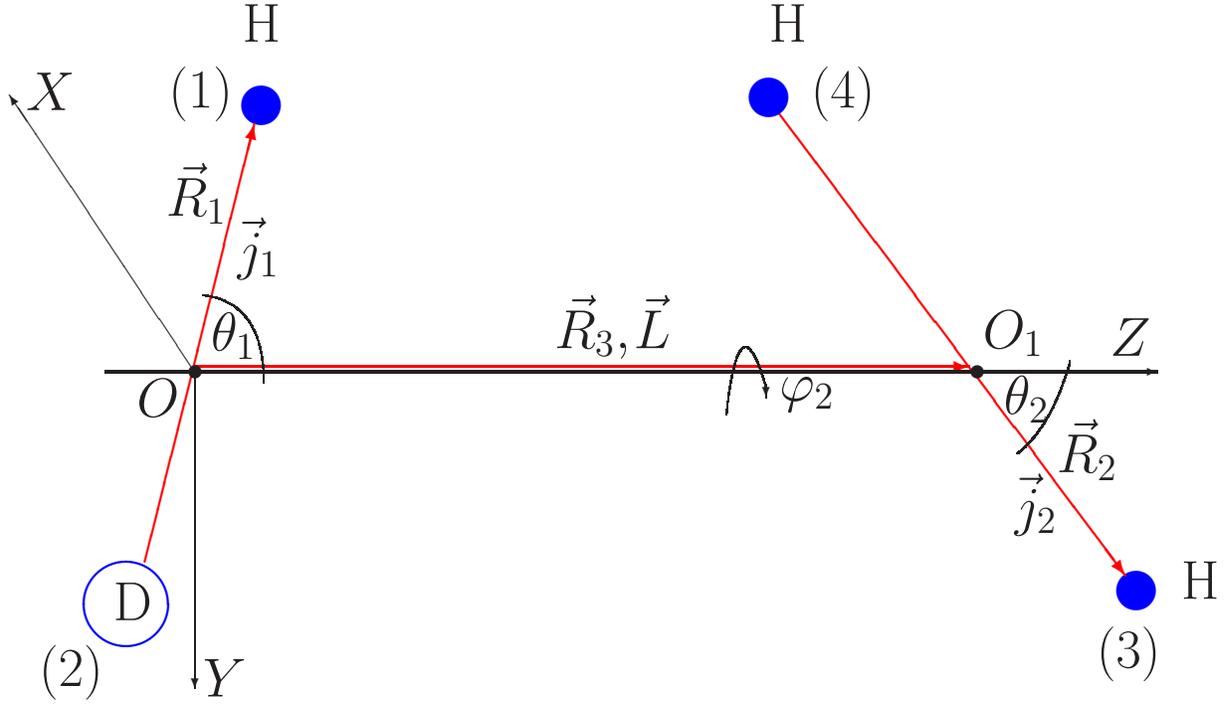
\begin{figure}
\begin{center}
{
\begin{picture}(500,250)(-80,0)

\thicklines{
{\color{blue}
\put(-28.5,23){\circle{32}}            
}}

{\color{blue}
\put(19,212){\circle*{23}}           
\put(350,28){\circle*{23}}            
}
\put(-65,-8){\LARGE {(2)}}      
\put(-16,211){\LARGE {(1)}}    
\put(227,211){\LARGE {(4)}}    
\put(335,-1){\LARGE {(3)}}    
\put(-37,17){\LARGE D}                               
\put(12,236){\LARGE H}                              
\put(211,236){\LARGE H}                             
\put(367,25){\LARGE H}                               
{\color{red}
\thicklines
\put(-25,39){\vector(1,4){41.5}}          

\put(214,210){\vector(3,-4){132}}     

{\color{blue}
\put(211,215){\circle*{23}}          
}

\put(-6,113){\vector(1,0){293}}      
}
\put(-6,111){\circle*{5}}        
\put(290,111){\circle*{5}}        
\put(292,120){\LARGE $O_1$}                      

\put(-40,111){\line(1,0){398}}       
\put(340,117){\LARGE $Z$}                      
\put(-70,210){\LARGE $X$}                       
\put(-3,-12){\LARGE $Y$}                          
\put(-28,94){\LARGE $O$}                      

\thinlines
\put(-40,111){\vector(1,0){398}}       
\put(-6.2,111){\vector(-2,3){70}}       
\put(-6.2,111){\vector(0,-1){120}}       

\put(300,95){\LARGE $\theta_2$}            

\qbezier(-3,140)(20,137)(20,107)   
\qbezier(325,115)(318,90)(305,80)   

\put(0,119){\LARGE $\theta_1$}              
\put(-17,170){\LARGE $\vec R_1$}             
\put(320,73){\LARGE $\vec R_2$}             
\put(304,53){\LARGE $\vec j_2$}             
\put(130,120){\LARGE $\vec R_3, $}          
\put(10,150){\LARGE $\vec j_1$}             
\put(160,120){\LARGE $\vec L$}              
\put(215,100){\LARGE $\varphi_2$}           
\qbezier(195,95)(200,140)(210,105)   
\put(210,105){\vector(0,-1){4}}
\end{picture}}
\end{center}
\vspace{10mm}
\caption{(Color online) Four-atomic system (12)+(34) or HD+H$_2$. Here, H is a hydrogen atom
and D is deuterium, represented by the few-body Jacobi coordinates: $\vec R_1$, $\vec R_2$ and
$\vec R_3$. The vector $\vec R_3$ connects the center of masses of the HD and H$_2$ molecules, i.e. $O$ and $O_1$
respectively, and is directed over the axis $OZ$, 
$\theta_1$ is the angle between $\vec R_1$ and $\vec R_3$,
$\theta_2$ is the angle between $\vec R_2$ and $\vec R_3$,
$\varphi_2$ is the torsional angle, $\vec j_1$, $\vec j_2$ and $\vec L$ are quantum angular
momenta over the corresponding Jacobi coordinates $\vec R_1$, $\vec R_2$ and $\vec R_3$.}
\label{fig:fig1}
\end{figure}

In this section we provide a brief account 
of the present quantum-mechanical close-coupling approach following 
the method in Ref. \cite{green75}.  
The HD and H$_2$ molecules are treated as linear rigid rotors. 
The model has been applied in few previous works  \cite{green75,flower99}.
In all our calculations with this potential the bond length was 
fixed at 1.449 a.u. or 0.7668 \r{A} for the H$_2$ molecule and
1.442 a.u. for HD which is 0.7631 \r{A}.
The Schr\"odinger equation for the  $(12)+(34)$ collision in the c.m.  frame,
where   $(12)$ and $(34)$ are 
diatomic molecules formed by atoms 1-4 is \cite{green75}:  
\begin{eqnarray}
\biggr[\frac{\hat{P}^2_{\vec R_3}}{2\mathcal{M}_{12}}&+&\frac{\hat{L}^2_{\hat R_{1}}}{2\mu_1R_1^2}+
\frac{\hat{L}^2_{\hat R_{2}}}{2\mu_2R_2^2}+
V(\vec R_1,\vec R_2,\vec R_3) - E \biggr]
\Psi(\hat R_{1},\hat R_{2},\vec R_3)=0.
\label{eq:schred}
\end{eqnarray}
Here $\hat{P}_{\vec R_3}$ is the momentum operator of the kinetic energy of
the collision, $\vec R_3$ is the collision 
coordinate, whereas $\vec R_1$ and $\vec R_2$ are relative vectors between 
atoms in the two diatomic molecules as shown in Fig.~\ref{fig:fig1},
and $\hat{L}_{\hat R_{1(2)}}$ are the quantum-mechanical
rotation operators of the rigid rotors, $\mu_1$ and $\mu_2$ are the reduced masses of the HD and H$_2$ 
molecules and $\mathcal{M}_{12}$ is the reduced mass of the two molecules.
The vectors $\hat R_{1(2)}$ are the angles of 
orientation for rotors $(12)$ and $(34)$, respectively;
$V(\vec R_1,\vec R_2,\vec R_3)$ is the PES of the four-atomic 
system $(1234)$, and $E$ is the total energy in the c.m.  system.
The use and modification of the original H$_2$-H$_2$ PESs
$V(\vec R_1,\vec R_2,\vec R_3)$ is discussed in Appendix A, i.e. Sec. 5.
The cross sections for rotational excitation and relaxation can be obtained 
from the $S$-matrix $S_{\alpha\alpha'}^J.$
The cross sections for excitation from HD($j_1$,$m_1$)+H$_2(j_2,m_2)$ to
HD$(j'_1m'_1)$+H$_2(j'_2m'_2)$ are summed over 
the final angular momentum projections ($m'_1m'_2$)
and averaged over the initial projections ($m_1m_2$)
of the HD and H$_2$ molecules of angular 
momenta $j_1$ and $j_2$ are given by:
\begin{eqnarray}
\sigma(j'_1,j'_2;j_1j_2,\varepsilon)=\frac{\pi}{(2j_1+1)(2j_2+1)k_{\alpha\alpha'}}  
\sum_{Jj_{12}j'_{12}LL'}(2J+1)|\delta_{\alpha\alpha'}-
S^J_{\alpha \alpha'}(E)|^2.
\label{eq:cross}
\end{eqnarray}
The kinetic energy is $\varepsilon=E-B_1j_1(j_1+1)-B_2j_2(j_2+1)$,
where $B_{1}=44.7\ \ \mbox{cm}^{-1}$ and
$B_{2}=60.8\ \ \mbox{cm}^{-1}$ are the rotation constants of rigid rotors $(12)$ and $(34)$ respectively,
they are shown in Fig.~\ref{fig:fig1}. Next, $E$ is the total energy of the system,
$J$ is total angular momenta of the four-atomic system, 
$\alpha \equiv (j_1j_2j_{12}L)$, where $j_1+j_2=j_{12}$ and $j_{12}+L=J$,
$k_{\alpha \alpha'}=\sqrt{2\mathcal{M}_{12}(E+E_{\alpha}-E_{\alpha'})}$ is the channel wavenumber and
$E_{\alpha(\alpha')}$ are rotational channel energies. Finally,
the relationship between the rotational thermal-rate coefficient 
$k_{j_1j_2\rightarrow j'_1j'_2}(T)$ at temperature $T$ and the corresponding
cross section $\sigma_{j_1j_2\rightarrow j'_1j'_2}(\varepsilon)$, can be obtained 
through the following weighted average formula:
\begin{equation}
k_{j_1j_2\rightarrow j'_1j'_2}(T) = \sqrt{ \frac{8k_B T}{\pi \mathcal{M}_{12}}}
\frac{1}{(k_BT)^2}
\int_{\varepsilon_s}^{\infty} \sigma_{j_1j_2\rightarrow j'_1j'_2}(\varepsilon)  
e^{-\varepsilon/k_BT}\varepsilon d\varepsilon,
\label{eq:kT}
\end{equation}
where $k_B$ is Boltzman constant  and $\varepsilon = E - E_{j_1} - E_{j_2}$
is pre-collisional translational energy at the
translational temperature $T$, $k_B$ is Boltzmann constant
and $\varepsilon_s$ is the minimum kinetic energy for the levels $j_1$ and $j_2$
to become accessible.

\section{Numerical  results}
\label{sec:results}
We used the MOLSCAT program \cite{hutson94} to solve the Schr\"odinger Eq.(\ref{eq:schred}).
Convergence was obtained for the integral cross sections, $\sigma(j'_1,j'_2;j_1j_2,\varepsilon)$,
with respect to the variation of  the variables utilized in all considered collisions at different collision
energies. For the intermolecular distance $R_3$ we used from $R_{3min}=3.0$ a.u.
to $R_{3max}=30.0$ a.u. We also applied a few different propagators included in the 
MOLSCAT computer program, and our calculation show that D. Manolopoulos's
hybrid modified log-derivative propagator technique \cite{manolop86}
would be quite numerically stable and a time effective approach. This method is
used in the majority of the calculations.

The maximum value of the total angular momentum $J$ was set at 44
while the number of levels $N_{lvl}$ in the basis set of HD + H$_2$ was set at 42.
Specifically, in the case of HD$(j_1)$+$para$-H$_2(j_2)$, $j_1$ it has values 0, 1, 2, and 3
and $j_2$ has values 0, 2, and 4. This combination generates the total number of included levels $N_{lvl}$=34.
In the case of HD+$ortho$-H$_2$ the parameter $j_1$ has values 0, 1, 2, and 3
and $j_2$ has values 1, 3, and 5. This results in the total number of levels $N_{lvl}$=42.
A number of test computations with higher values for the $j_1$ and $j_2$
parameters have been carried out. For example, in the case of HD+$para$-H$_2$, $j_1$
was taken as 0, 1, 2, 3, 4 and $j_2$ as 0, 2, 4, 6. This  $j_1$/$j_2$ combination
produced $N_{lvl}=74$. We obtained similar results in both cases, confirming 
the convergence of the calculations.

Because the HD+H$_2$ total rotational energy transfer cross sections (\ref{eq:cross})
has shape resonances at low energies a large number of energy
points are needed in order to effectively reproduce them. 
We used up to 250 energy points in each computation
for each specific rotational transition in the HD and H$_2$ molecules considered.
More space discretization points were used at low collision
energies and fewer points in the  higher-energy sector.  

\begin{figure}\begin{center}            
\includegraphics*[scale=1.0,width=21pc,height=13pc]{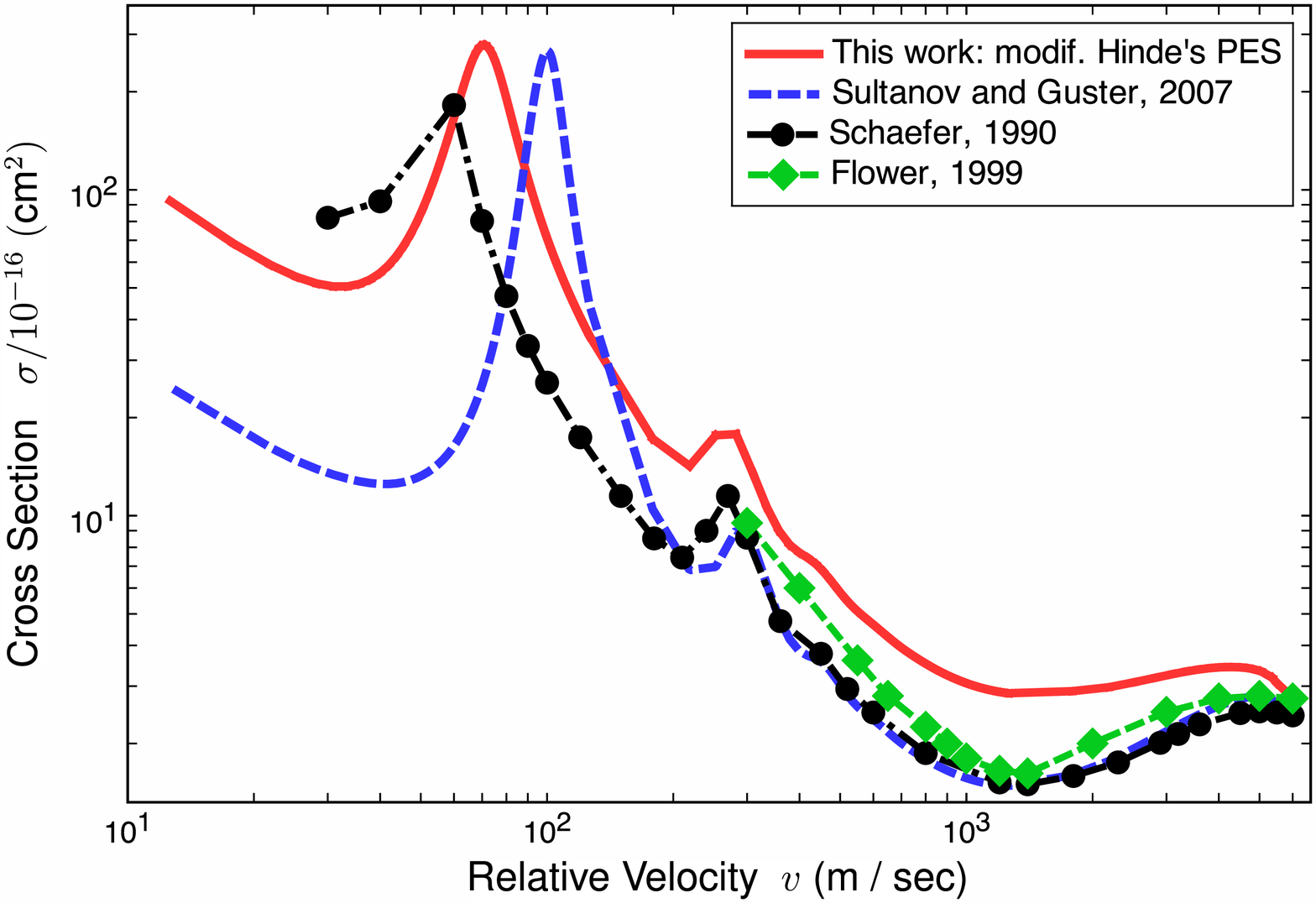}
\caption{(Color online) The total cross section of the HD(1) + H$_2$(0) $\rightarrow$ HD(0) + H$_2$(0)
inelastic rotational energy transfer collision. The numbers in the brackets are the rotational
quantum numbers of the corresponding two-atomic molecules.}\label{fig:fig3}\end{center}\end{figure}

\begin{figure}\begin{center}            
\includegraphics*[scale=1.0,width=21pc,height=13pc]{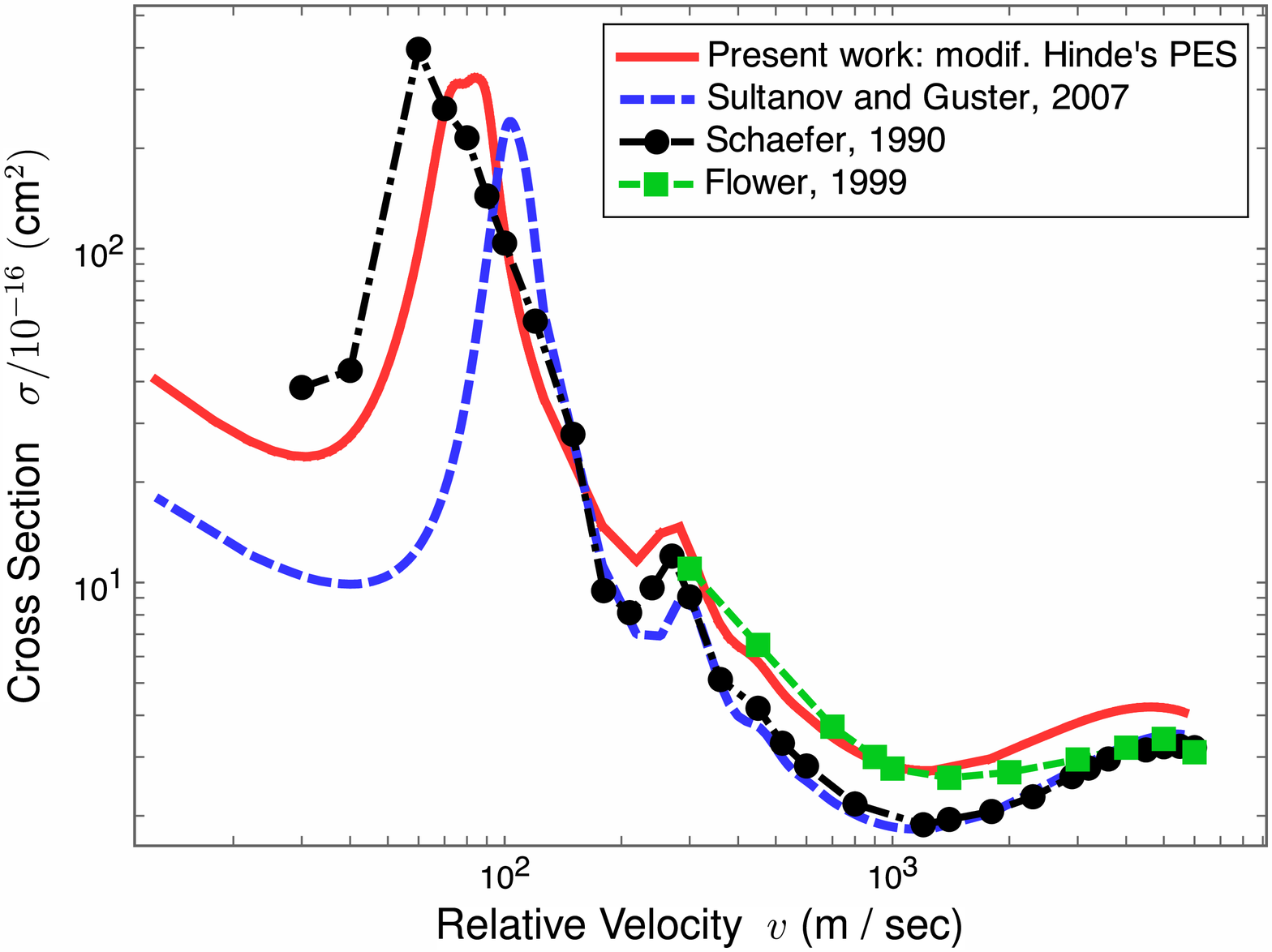}
\caption{(Color online) The total cross section of the HD(2) + H$_2$(0) $\rightarrow$ HD(1) + H$_2$(0)
inelastic rotational energy transfer collision. The numbers in the brackets are the rotational
quantum numbers of the corresponding two-atomic molecules.}\label{fig:fig4}\end{center}\end{figure}

\begin{figure}\begin{center}            
\vspace{-2mm}
\includegraphics*[scale=1.0,width=21pc,height=13pc]{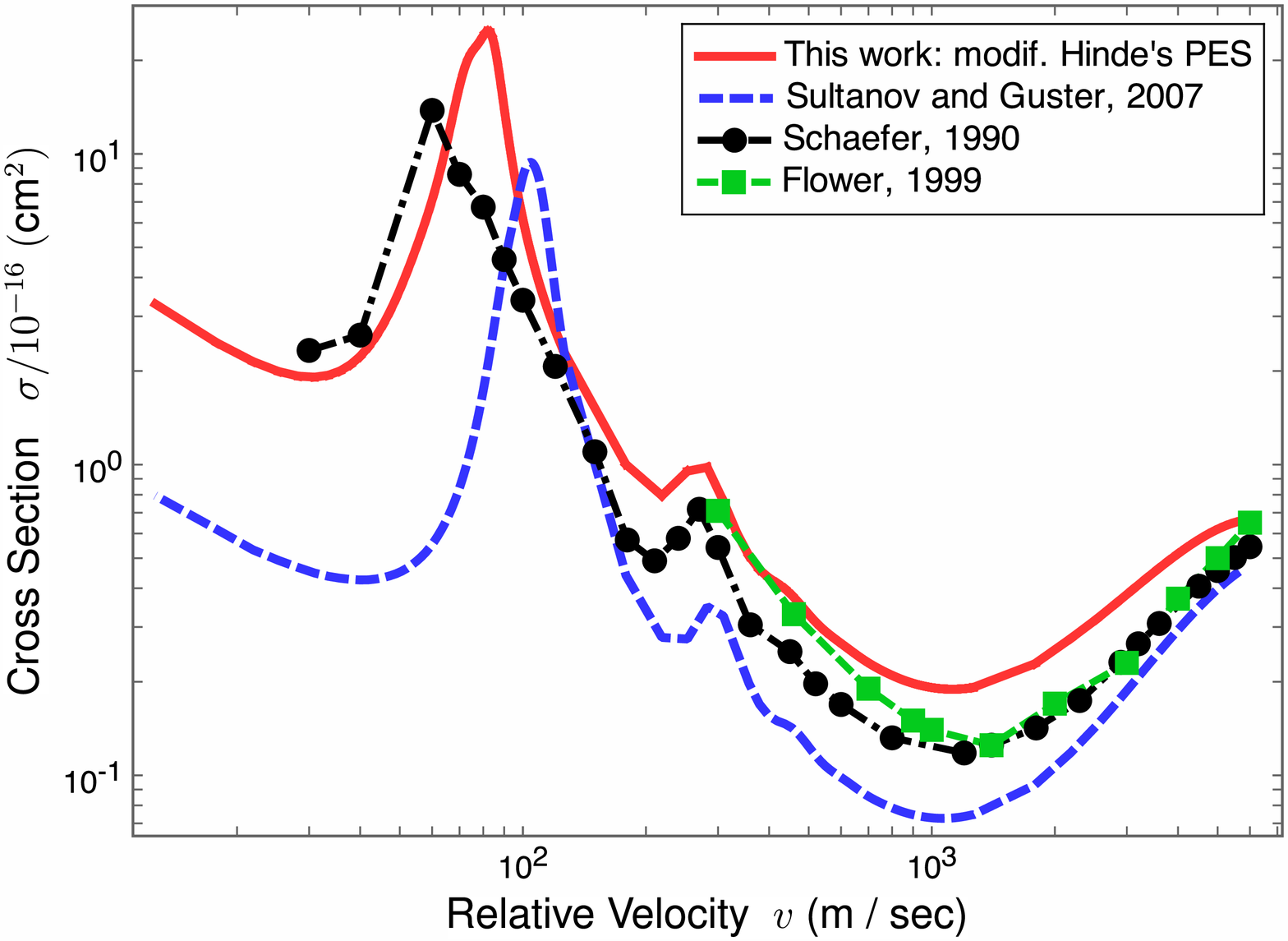}
\caption{(Color online) The total cross section of the HD(2) + H$_2$(0) $\rightarrow$ HD(0) + H$_2$(0)
inelastic rotational energy transfer collision. The numbers in the brackets are the rotational
quantum numbers of the corresponding two-atomic molecules.}\label{fig:fig5}\end{center}\end{figure}

\begin{figure}
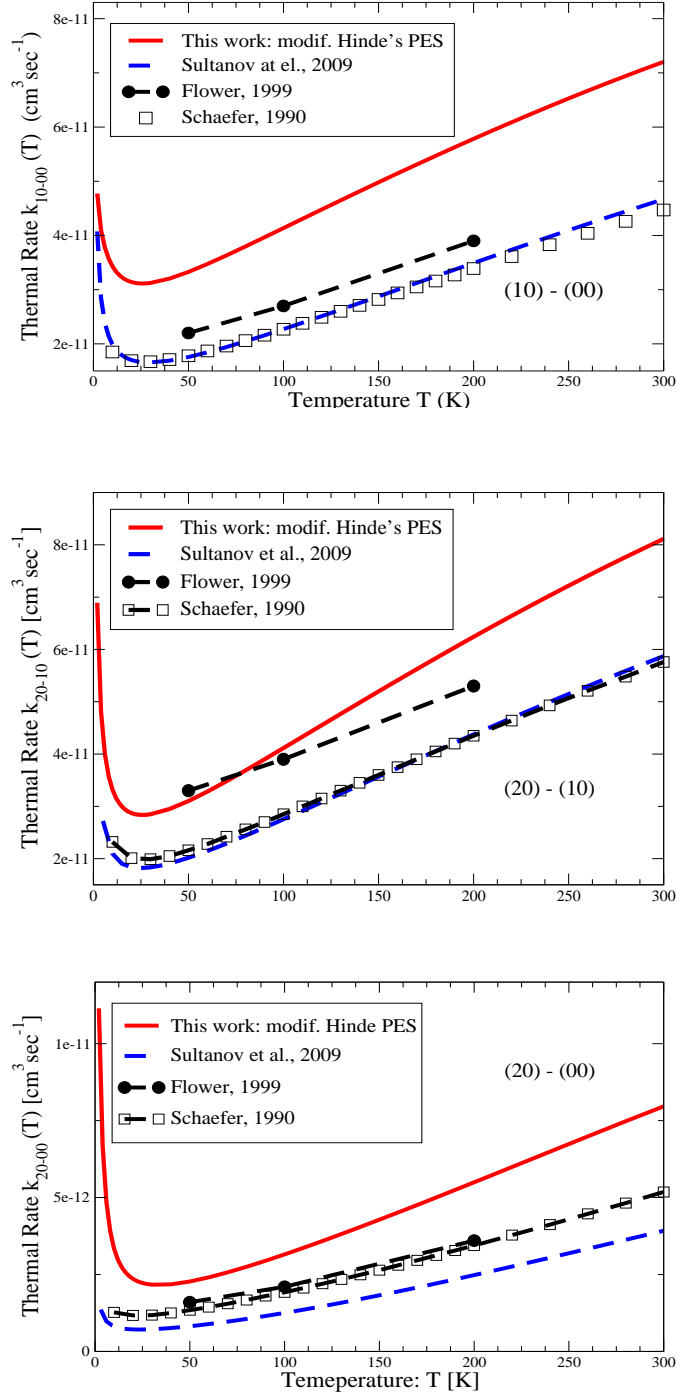
\begin{center}            
\includegraphics[scale=1.0,width=21pc,height=13pc]{Fig913_OSU_10-00_NEW_RH.eps}\vspace{10mm}\\
\includegraphics[scale=1.0,width=21pc,height=13pc]{Fig911_OSU_20-10_NEW_RH.eps}\vspace{10mm}\\
\includegraphics[scale=1.0,width=21pc,height=13pc]{Fig912_OSU_20-00_NEW_RH.eps}\vspace{5mm}\\
\caption{(Color online) Upper plot: the rotational de-excitation thermal rate coefficients $k_{ij\rightarrow i'j'}(T)$ for
the $\mbox{HD}(1) + \mbox{H}_2(0) \rightarrow \mbox{HD}(0) + \mbox{H}_2(0)$ collision.
Middle and lower plots represent $k_{ij\rightarrow i'j'}(T)$ for the 
$\mbox{HD}(2) + \mbox{H}_2(0) \rightarrow \mbox{HD}(1) + \mbox{H}_2(0)$ and
$\mbox{HD}(2) + \mbox{H}_2(0) \rightarrow \mbox{HD}(0) + \mbox{H}_2(0)$ channels
correspondingly.}\label{fig:fig6}\end{center}\end{figure}

Below we present results for cross sections and thermal rate coefficients for different 
quantum-state transitions in HD and H$_2$ molecules. 
We reproduced shape resonances in the low velocity region, which are 
very important in the cooling of the astrophysical media at low temperatures. 
We compared them with the older quantum-dynamical results
of Schaefer \cite{schaefer} and Flower \cite{flower99} .  
We also present results \cite{my1,my2} using a PES 
obtained from a modification of the BMKP H$_2$-H$_2$ PES 
\cite{boothroyd}, viz. Sec.  \ref{sec:HD_PES}  (Appendix A).
In Figs. \ref{fig:fig3},\ref{fig:fig4} and \ref{fig:fig5} we show 
four different results for the integral cross sections in the collisions:
\begin{eqnarray}\label{eq:1000}
\rm{HD}(1) + H_2(0) \rightarrow HD(0) + H_2(0), \\
\label{eq:2010}
\rm{HD}(2) + H_2(0) \rightarrow HD(1) + H_2(0), \\
\label{eq:2000}
\rm{HD}(2) + H_2(0) \rightarrow HD(0) + H_2(0), 
\end{eqnarray}
respectively. 
One can see that the new cross sections obtained
with the present  PES  have the same structure and shape, but also have
substantially larger ($\sim$ 60\%) values at medium energies ($v>200$ m/sec) when
compared with the  results obtained from the modified
BMKP PES \cite{boothroyd} and the older result from Schaefer's calculations \cite{schaefer}.
Processes (\ref{eq:2010}) and (\ref{eq:2000}) are also important
collisions from the astrophysical point of view. 
They represent transitions from the $j_1=2$ and $j_2=0$ state of the 
HD$(j_1)$+H$_2(j_2)$ system.
It is seen from Table \ref{tab:tab1} that the corresponding
rotational energy of the system is 268.2 cm$^{-1}$ and rotational 
relaxation processes to states of lower energy should be relevant.
The cross sections from Ref.  \cite{flower99} are available only
for higher collision velocities, i.e. $v \geq$ 300 m/s. 
The  present  cross sections have the 
same behavior as previous results, but  they 
have larger values than the old ones, specially at higher energies.
Next  in Fig. \ref{fig:fig6}
we exhibit  the thermal rate coefficients corresponding  to the
processes considered above.
Here, we additionally include the older results from Ref. \cite{flower99},
where Schwenke's H$_2$-H$_2$ PES \cite{schwenke88}
was used. One can see that the new rates are substantially larger than those of other calculations.
In general, it means that the contribution of the HD+H$_2$ collision to the HD-cooling function can have
even larger contributions than previously expected.

\begin{figure}\begin{center}            
\includegraphics*[scale=1.0,width=21pc,height=13pc]{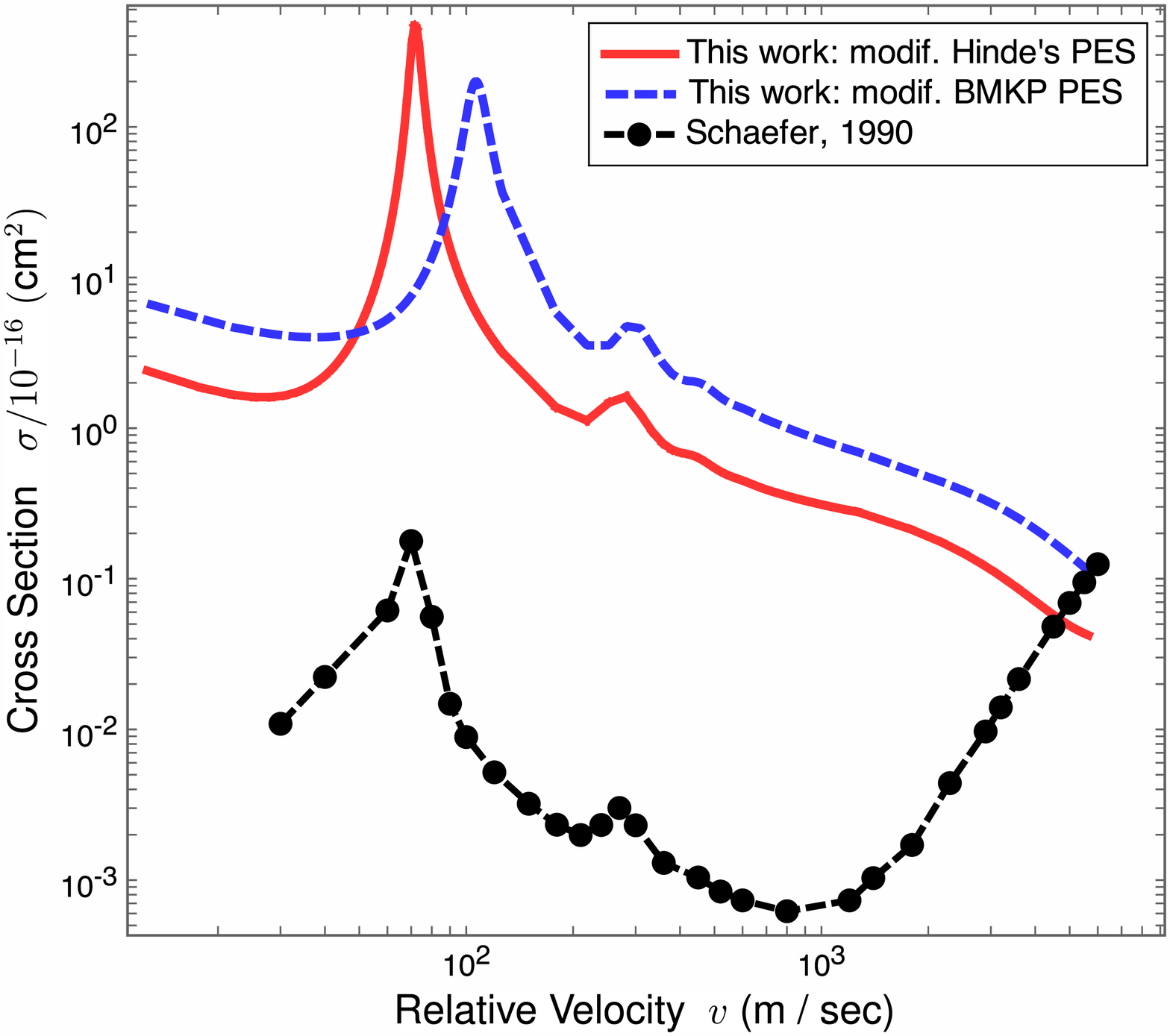}\vspace{5mm}\\
\includegraphics*[scale=1.0,width=21pc,height=13pc]{Fig7_02-20_kT_all.eps}
\caption{(Color online) Upper plot: 
the total cross section of the HD(0) + H$_2$(2) $\rightarrow$ HD(2) + H$_2$(0)
inelastic rotational energy transfer collision. The numbers in the brackets are the rotational
quantum numbers of the corresponding two-atomic molecules.
Lower plot: the corresponding thermal rate coefficient, i.e. the process (02)-(20).}
\label{fig:fig7}\end{center}\end{figure}

Now we consider the inelastic cross sections in HD+H$_2$ for higher rotational
energies. For the systems 
HD(0)+H$_2(2)$ and HD(1)+H$_2(2)$, one can see from Table \ref{tab:tab1}
that  the corresponding rotational energies are
364.8 cm$^{-1}$ and 454.2 cm$^{-1}$. In Fig.~\ref{fig:fig7} (upper plot)
we  show the integral cross sections for the process:
\begin{equation}\label{eq:0220}
\rm{HD}(0) + H_2(2) \rightarrow HD(2) + H_2(0). 
\end{equation}
While we obtain a relatively good agreement between
cross sections calculated
with the older modified BMKP PES and Hinde's PES, there is a dramatic difference  with
the corresponding result from Ref.  \cite{schaefer}. The present cross sections are larger than those of Ref.  \cite{schaefer}
by a few orders of magnitude. 
In Fig.~\ref{fig:fig7} (lower plot) we show the results for the corresponding thermal rate coefficient $k_{20\rightarrow 02}(T)$.
It is seen, that the present rates and those obtained by Flower \cite{flower99}  have a flat
temperature dependence, whereas the rate of Schaefer \cite{schaefer}, 
although smaller than other results,  increases monotonically with energy. 
Because the thermal rate of reaction (\ref{eq:0220}) is relatively large,
one can conclude that this channel can also make a substantial 
contribution to the astrophysical HD-cooling function.

In Figs. \ref{fig:fig8}-\ref{fig:fig11} we  show results for the total cross section in the inelastic scattering from
the state $\rm{HD}(1) + H_2(2)$ with  
 rotational energy  454.2 cm$^{-1},$ viz. Table~\ref{tab:tab1}. All de-excitation processes have been
computed for this state, namely: 
\begin{eqnarray}
\rm{HD}(1) + H_2(2) \rightarrow HD(0) + H_2(2),\label{eq:1202}\\
\rm{HD}(1) + H_2(2) \rightarrow HD(2) + H_2(0),\label{eq:1220}\\
\rm{HD}(1) + H_2(2) \rightarrow HD(1) + H_2(0),\label{eq:1210}\\
\rm{HD}(1) + H_2(2) \rightarrow HD(0) + H_2(0).\label{eq:1200}
\end{eqnarray}
The present cross sections exibit fairly good agreement with the
results computed with the modified BMKP PES \cite{boothroyd},
and also with older results from Ref.  \cite{schaefer}. The general shape
and trend of the behavior of these cross sections are the same in all cases. 
 Also, there is a relatively small bump in the cross sections of the processes
(\ref{eq:1220})-(\ref{eq:1200}) at collision velocity $\sim$1100 m/sec which is also reproduced by two PESs.
One can see that the process (\ref{eq:1202}) can make a significant contribution to the total HD-cooling function because
its cross section is rather large relative to other channels.
Finally, in order to carry out new computations of the astrophysical cooling function, Table~\ref{tab:tab2} 
includes the relevant thermal rate coefficients for the HD + $para$-H$_2$ case in the temperature range
from 2 K to 300 K.
Next, as a test, we choose the initial state HD(2) + H$_2$(2)
with a relatively higher total rotational energy: 633 cm$^{-1}$.
The integral de-excitation cross sections from this state to different 
lower energy rotational states are shown in Fig.~\ref{fig:fig12}.
One can see a fairly good agreement in the shape of the curves
between various rotational transition results for the integral cross sections
computed with the two different PESs.

In Fig.~\ref{fig:fig13} we show three different rotational transition
cross-sections for the $ortho$-hydrogen case.
Here we chose low lying rotational levels of the two molecules:
HD(1) + H$_2$(1) and HD(2) + H$_2$(1). Some
results from older works \cite{my22, schaefer,flower99} are also presented in the figure
together with results computed with the newer modified PES from \cite{hinde}.
The corresponding thermal rate coefficients are shown
in Fig.~\ref{fig:fig14} where the older 
results from Refs. \cite{my22,schaefer,flower99} are also presented
for comparison purposes.
Figures ~\ref{fig:fig15} and \ref{fig:fig16} include results for
thermal rate coefficients for the transitions from higher rotational states, e.g.,
HD(0) + H$_2$(3) and HD(1) + H$_2$(3). 

Finally, in Sec. \ref{sec:rates7} (Appendix B)
we present our new results for the thermal rate coefficients which can be used in
subsequent computation of the astrophysical HD-cooling function.
In Table~\ref{tab:tab2} we show thermal rates of different 
de-excitation processes in low-temperature HD + $para$-H$_2$ 
rotational energy transfer collisions and in Table~\ref{tab:tab3} the same
data for  the HD + $ortho$-H$_2$ case. 

\section{Summary and conclusions}
\label{sec:conclusion}
State-to-state close-coupling quantum-mechanical calculations for rotational de-excitation
cross-sections and corresponding thermal rate coefficients of the HD+$o$-/$p$-H$_2$ 
collisions are presented using a linear rigid rotor model for the HD and H$_2$ molecules.
The  symmetrical H$_2$-H$_2$ PES of Ref. \cite{hinde} has been appropriately adopted for
the current non-symmetrical HD+H$_2$ system by appropriate translation and rotation. 
These geometrical operations lead to  a new set of angle variables 
$\theta^{\prime}_1, \theta^{\prime}_2$ and $\varphi^{\prime}_2$
for the Jacobi few-body coordinates, a new length of the intermolecular distance 
$\vec R^{\prime}_3$ and, as a result, to a new HD-H$_2$ PES.
For comparison purposes in this paper we carried out a few calculations with the use of
the older BMKP PES\cite{boothroyd} which was also modified for HD-H$_2$. 
A test of convergence and the results for the cross-sections with 
the two PESs are obtained for a wide range of values of different parameters.
It is seen from Figs. \ref{fig:fig3}, \ref{fig:fig4} and \ref{fig:fig5}, 
that in most cases for the lower number quantum transition states 
the rotational energy transfer cross sections obtained  with the use of the modified Hinde's PES
have higher values than the cross sections computed with the use of the modified BMKP PES\cite{my2,my22}.
The same situation occurs in comparisons within the older work \cite{schaefer}.
In Fig. \ref{fig:fig6} the corresponding thermal rate coefficients
are presented. In this case we  include results from the paper\cite{flower99} too. 
As can be seen the new thermal rates (solid lines) have
substantially higher values than the other results. For example, in the rotational energy transfer, process (\ref{eq:1000}),
new thermal rates are $\sim$2 times higher than the other corresponding rates. Because (\ref{eq:1000}) is considered as one of the
main contributors to the resulting HD cooling process, one can say that the new HD-cooling function may have 
substantially greater values than the previous calculations\cite{flower1}. 
Next, Fig. \ref{fig:fig7} represents 
results for a very interesting process (\ref{eq:0220}). The interest in this channel lies in the fact that both molecules change their
rotational quantum numbers by $\Delta j$=2, therewith HD becomes excited and H$_2$ de-excited. In this case we obtained a
significant deviation from the results of work\cite{schaefer}. However, 
the cross sections obtained with the modified versions of the Hinde and 
BMKP PESs are fairly close to each other and have fairly large values. Therefore, this process probably can make a
contribution to the HD cooling process. In addition we would like to note, that in work\cite{flower1} the process has also been
computed and discussed, in which significant differences from\cite{schaefer} were also found.

\begin{figure}\begin{center}            
\includegraphics*[scale=1.0,width=21pc,height=13pc]{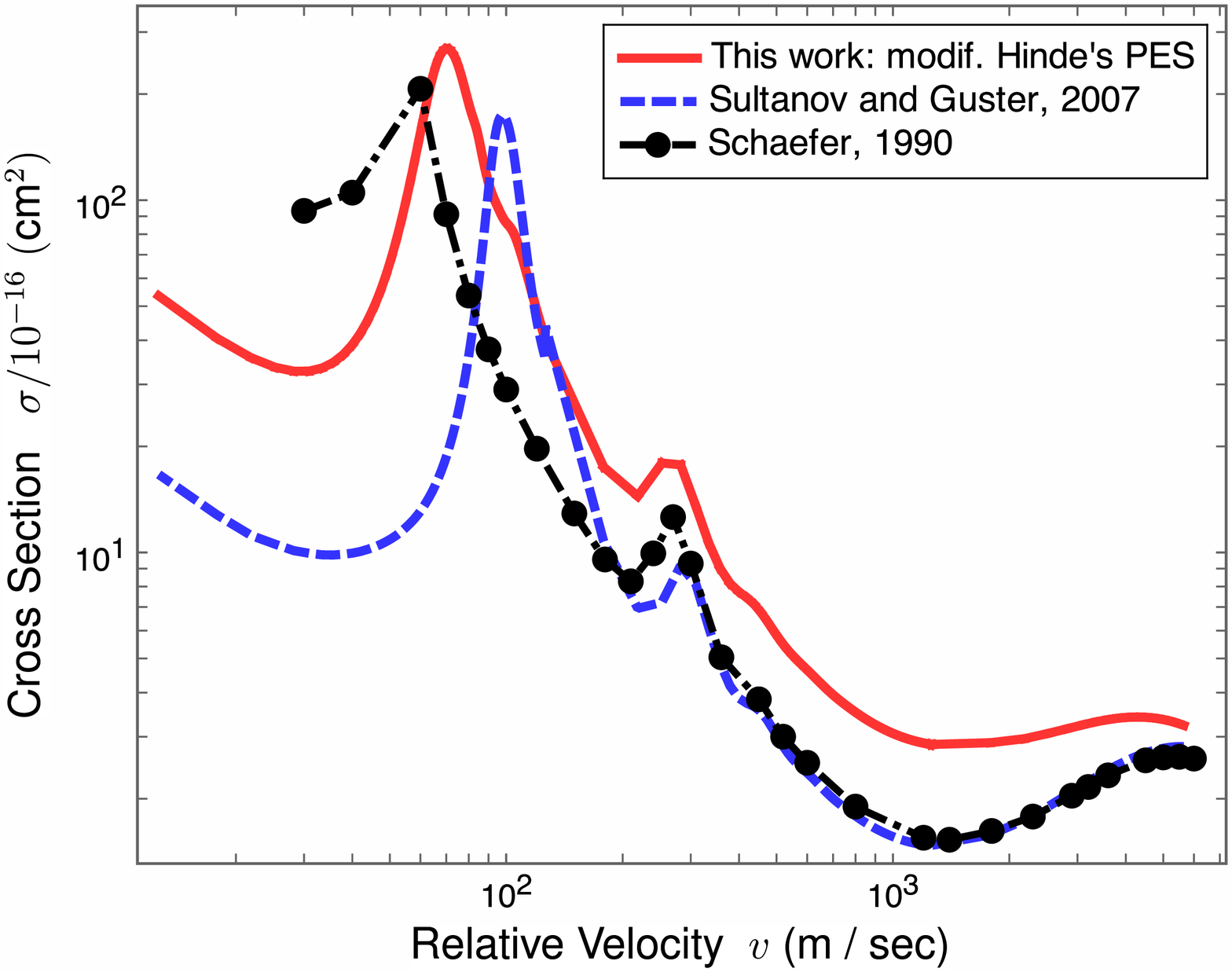}
\caption{(Color online) The total cross section (upper plot) and thermal rate coefficients (lower plot)
of the HD(1) + H$_2$(2) $\rightarrow$ HD(0) + H$_2$(2)
inelastic rotational energy transfer collision. The numbers in the brackets are the rotational
quantum numbers of the corresponding two-atomic molecules.}\label{fig:fig8}\end{center}\end{figure}

\begin{figure}\begin{center}            
\includegraphics*[scale=1.0,width=21pc,height=13pc]{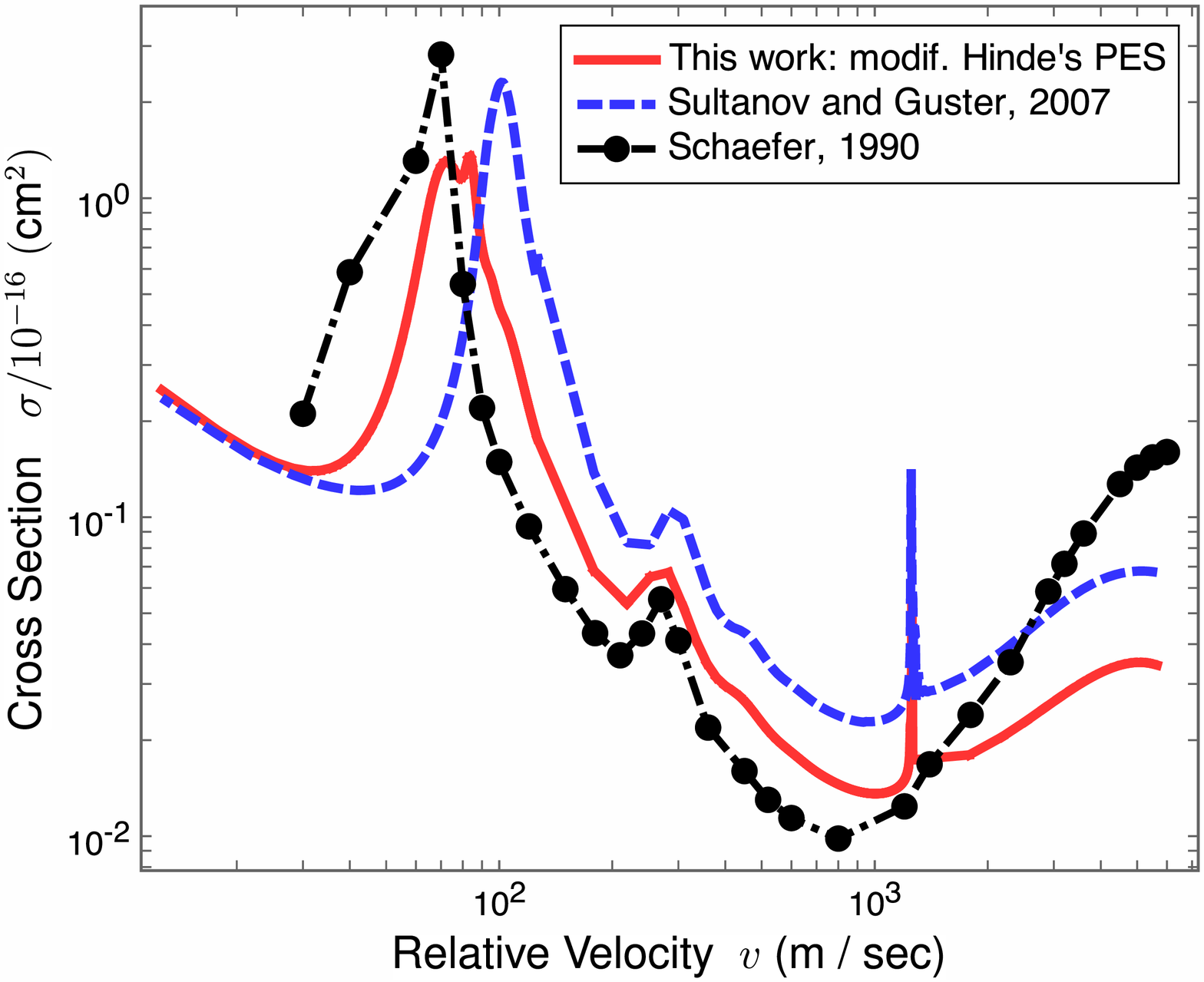}
\caption{(Color online) The total cross section of the HD(1) + H$_2$(2) $\rightarrow$ HD(2) + H$_2$(0)
inelastic rotational energy transfer collision. The numbers in the brackets are the rotational
quantum numbers of the corresponding two-atomic molecules.}\label{fig:fig9}\end{center}\end{figure}

\begin{figure}\begin{center}            
\includegraphics*[scale=1.0,width=21pc,height=13pc]{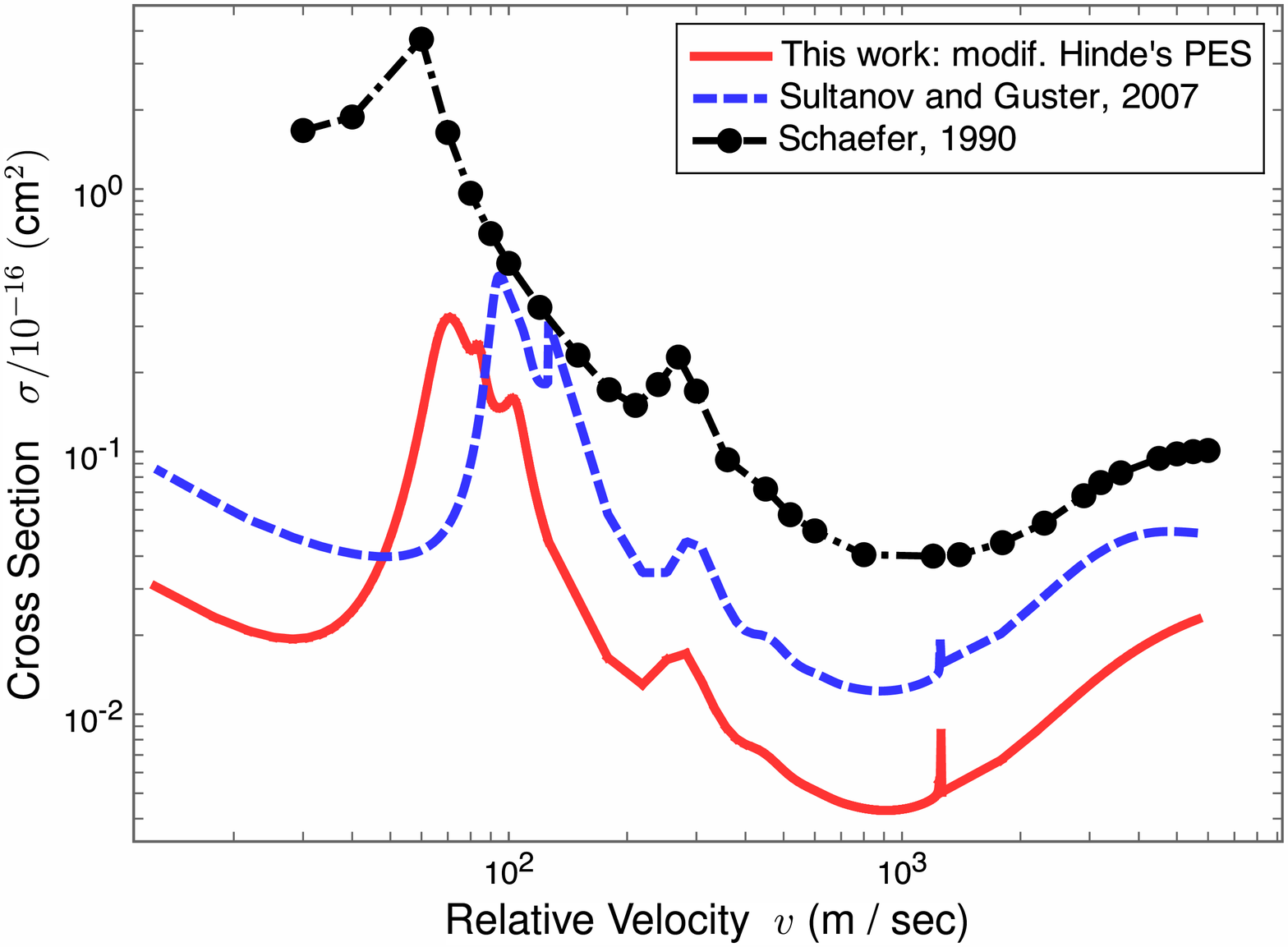}
\caption{(Color online) The total cross section of the HD(1) + H$_2$(2) $\rightarrow$ HD(1) + H$_2$(0)
inelastic rotational energy transfer collision. The numbers in the brackets are the rotational
quantum numbers of the corresponding two-atomic molecules.}\label{fig:fig10}\end{center}\end{figure}

\begin{figure}\begin{center}            
\includegraphics*[scale=1.0,width=21pc,height=13pc]{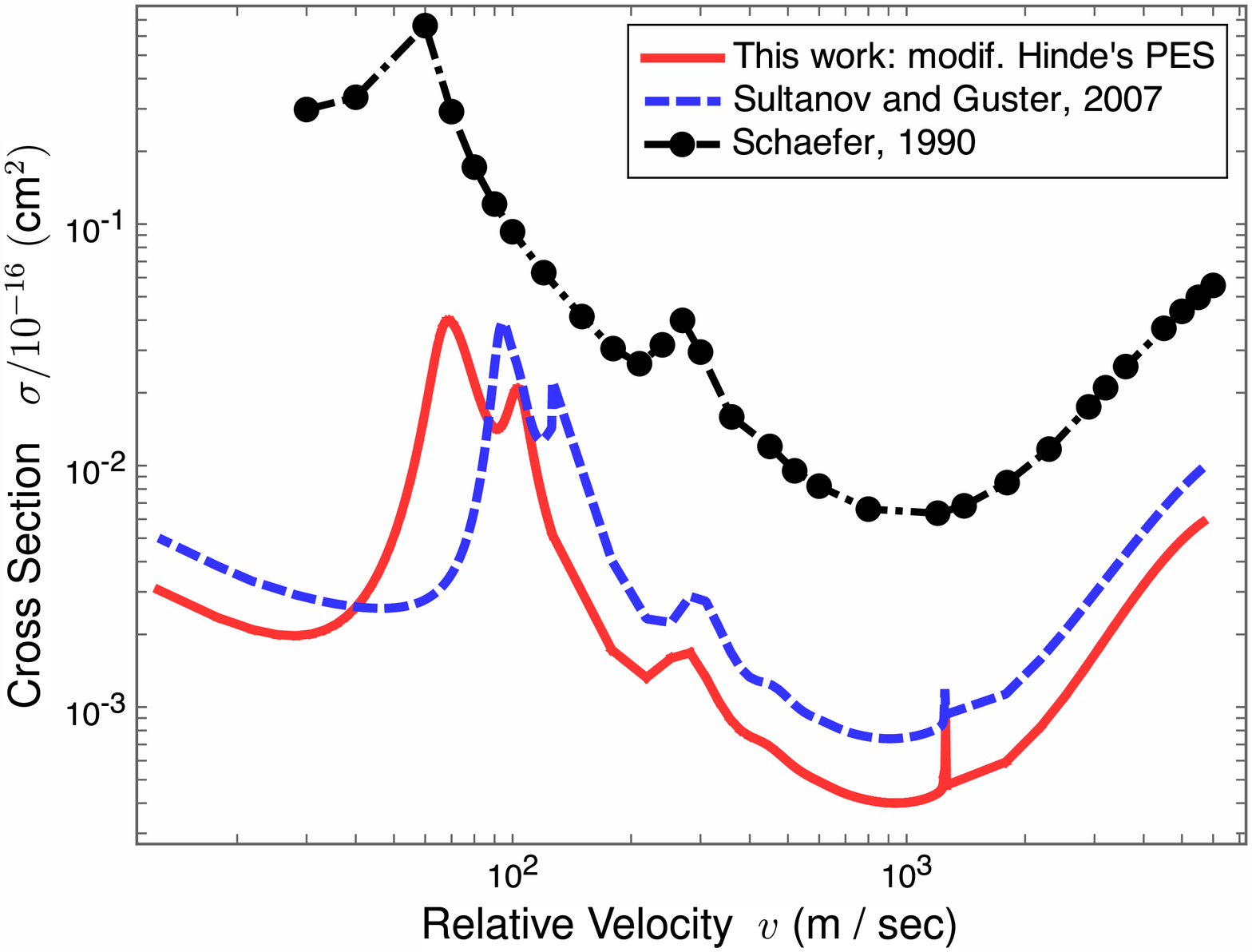}
\caption{(Color online) The total cross section of the HD(1) + H$_2$(2) $\rightarrow$ HD(0) + H$_2$(0)
inelastic rotational energy transfer collision. The numbers in the brackets are the rotational
quantum numbers of the corresponding two-atomic molecules.}\label{fig:fig11}\end{center}\end{figure}

\begin{figure}\begin{center}            
\includegraphics*[scale=1.0,width=21pc,height=13pc]{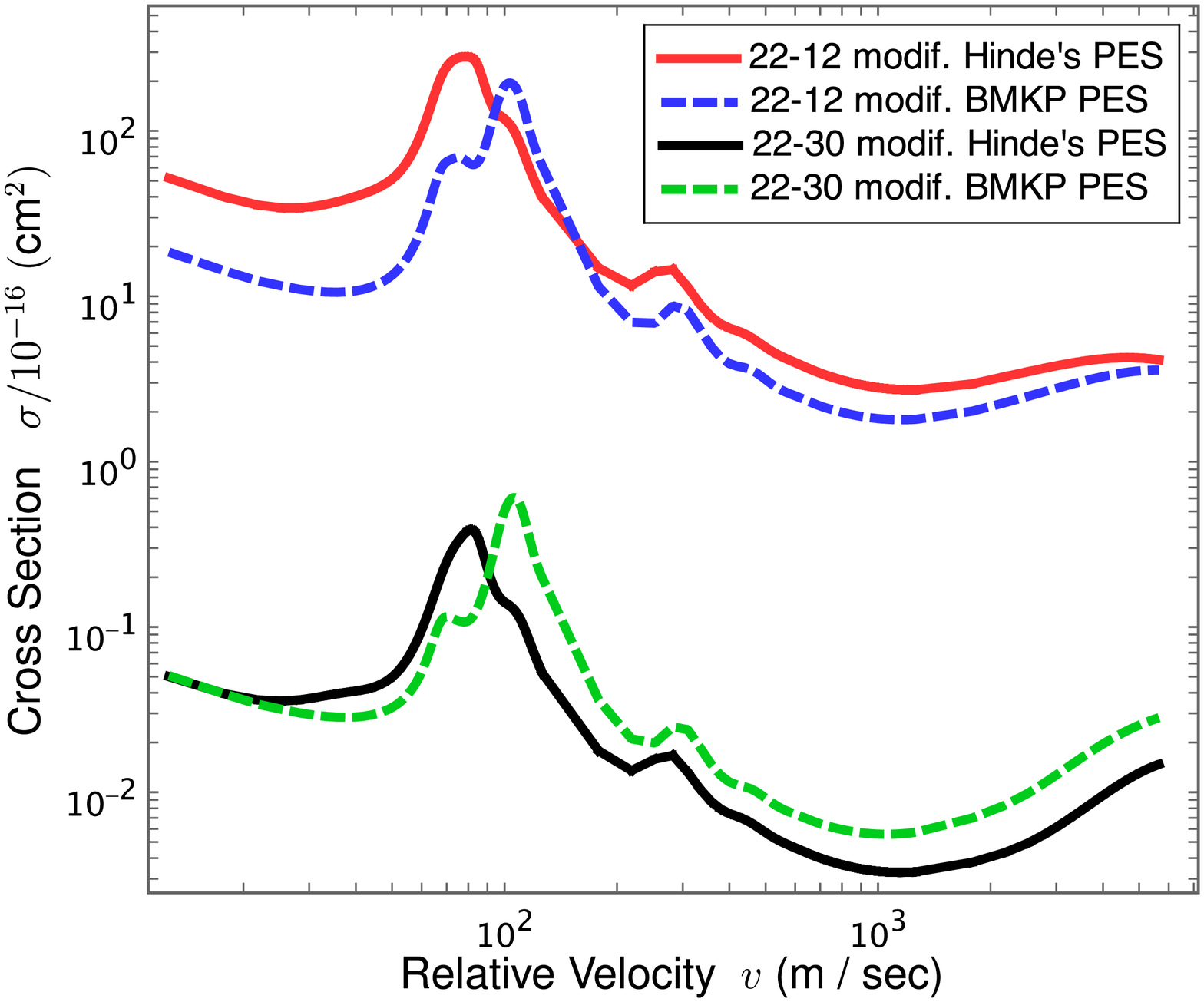}\vspace{7mm}\\
\includegraphics*[scale=1.0,width=21pc,height=13pc]{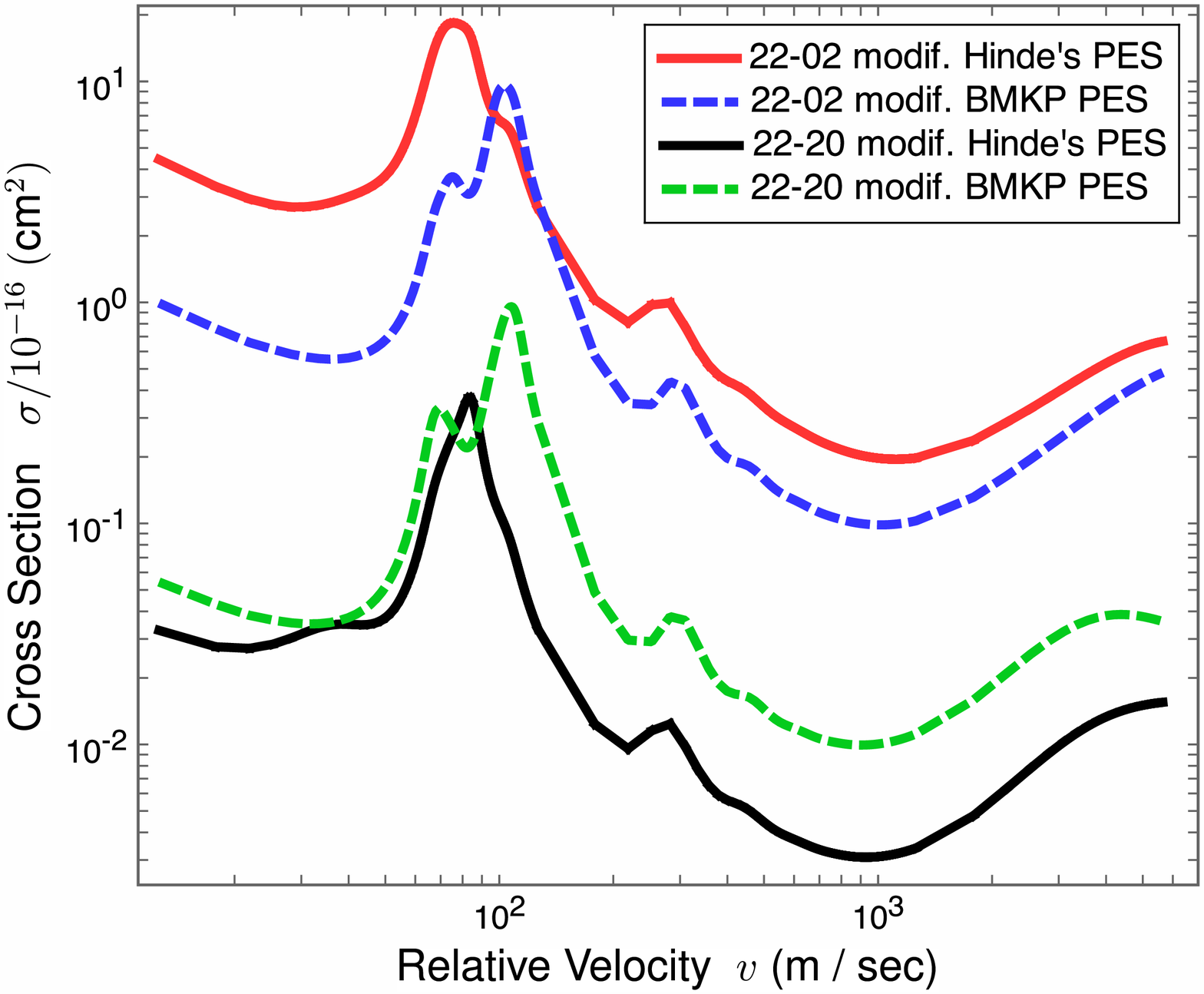}
\caption{(Color online)
Upper graph: the total cross sections of the HD(2) + H$_2$(2) $\rightarrow$ HD(1) + H$_2$(2)
and HD(2) + H$_2$(2) $\rightarrow$ HD(3) + H$_2$(0)
inelastic rotational energy transfer collisions. 
Lower graph: the total cross sections of the HD(2) + H$_2$(2) $\rightarrow$ HD(0) + H$_2$(2)
and HD(2) + H$_2$(2) $\rightarrow$ HD(2) + H$_2$(0)
inelastic rotational energy transfer collisions. The numbers in the brackets are the rotational
quantum numbers of the corresponding two-atomic molecules. Results are obtained with
two different PESs.}\label{fig:fig12}\end{center}\end{figure}

\begin{figure}
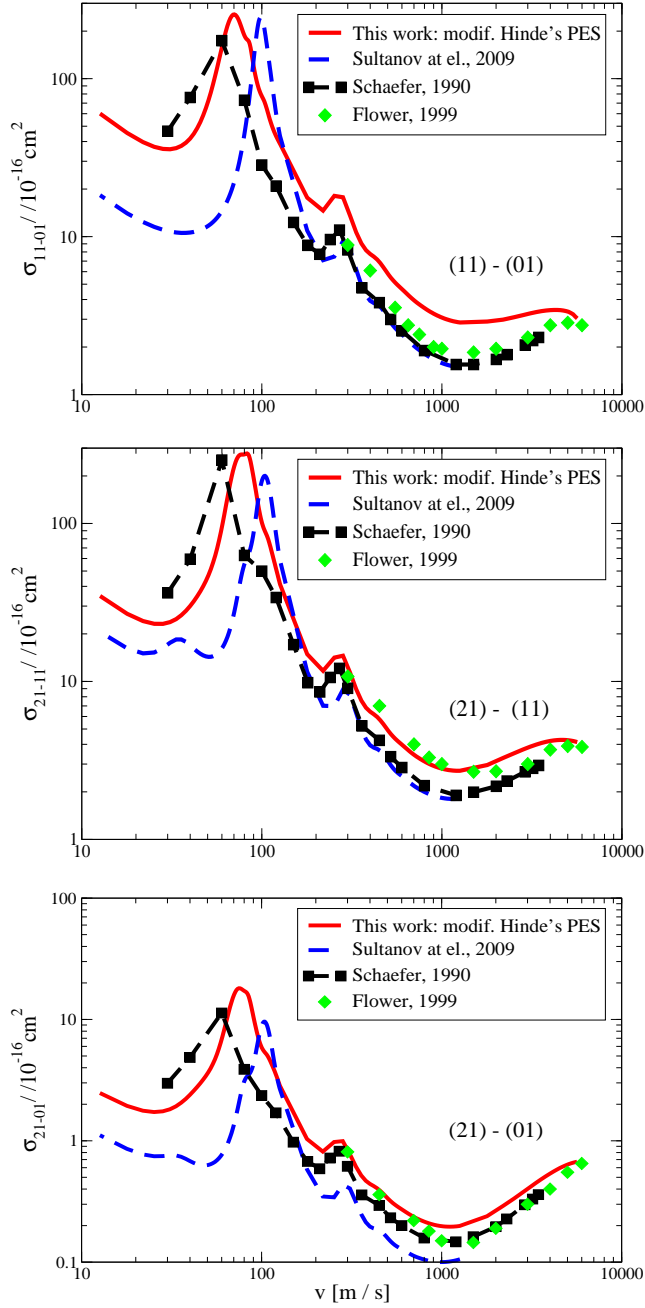
\begin{center}            
\includegraphics*[scale=1.0,width=20pc,height=13pc]{Fig8OSU_1101_NEW_RH.eps}\vspace{4mm}\\
\includegraphics*[scale=1.0,width=20pc,height=13pc]{Fig91OSU_2111_NEW_RH.eps}\vspace{4mm}\\
\includegraphics*[scale=1.0,width=20pc,height=13pc]{Fig9OSU_2101_NEW_RH.eps}
\caption{(Color online) The HD+$ortho$-H$_2$ case.
Upper plot: the total cross section of the HD(1) + H$_2$(1) $\rightarrow$ HD(0) + H$_2$(1)
inelastic rotational energy transfer collision. Middle and lower plots represent the following two
integral cross sections:
HD(2) + H$_2$(1) $\rightarrow$ HD(1) + H$_2$(1) and HD(2) + H$_2$(1) $\rightarrow$ HD(0) + H$_2$(1)
correspondingly. The numbers in the brackets are the rotational
quantum numbers of the corresponding two-atomic molecules.}\label{fig:fig13}\end{center}\end{figure}

\begin{figure}\begin{center}            
\includegraphics*[scale=1.0,width=20pc,height=13pc]{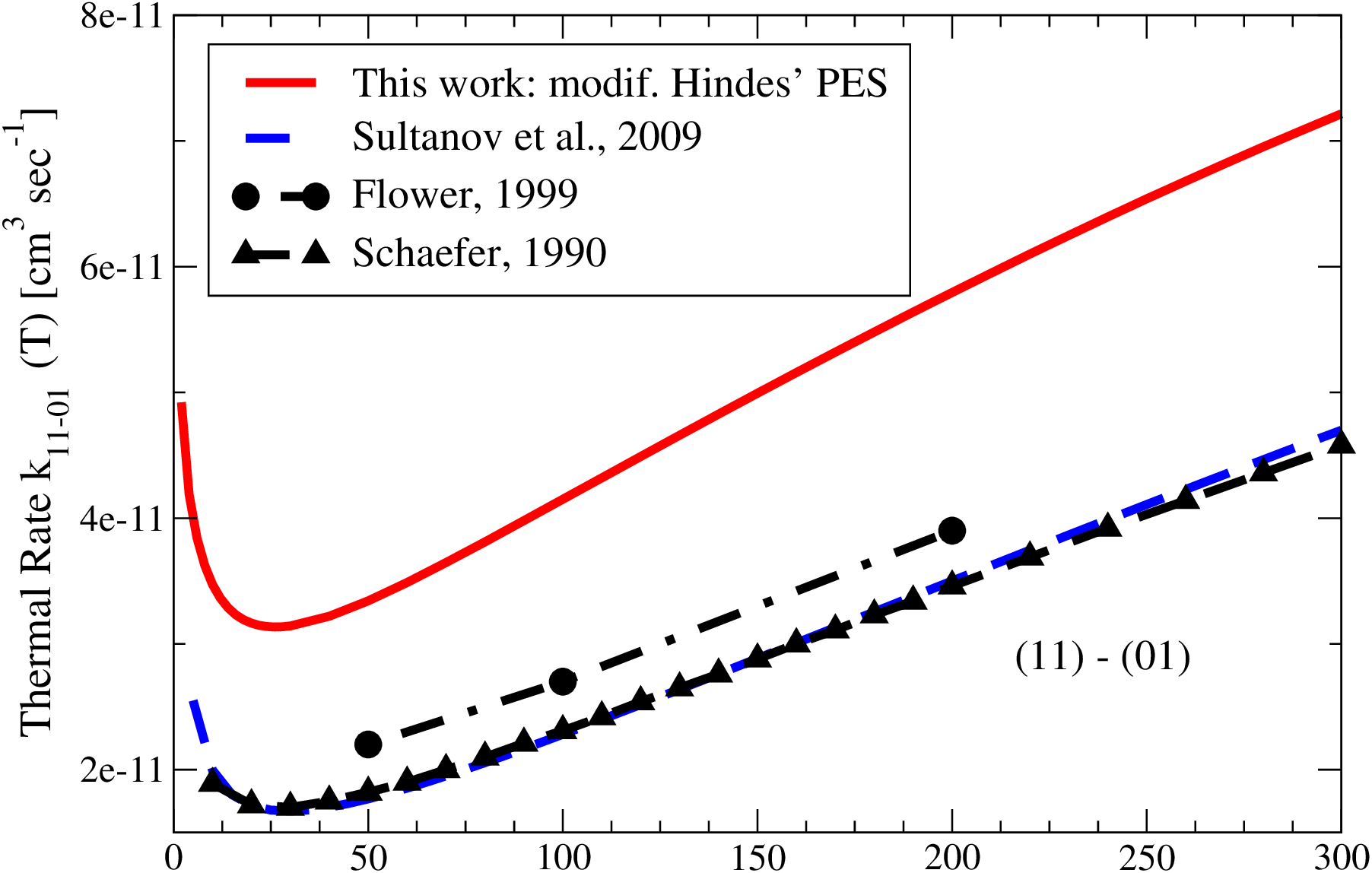}\vspace{4mm}\\
\includegraphics*[scale=1.0,width=20pc,height=13pc]{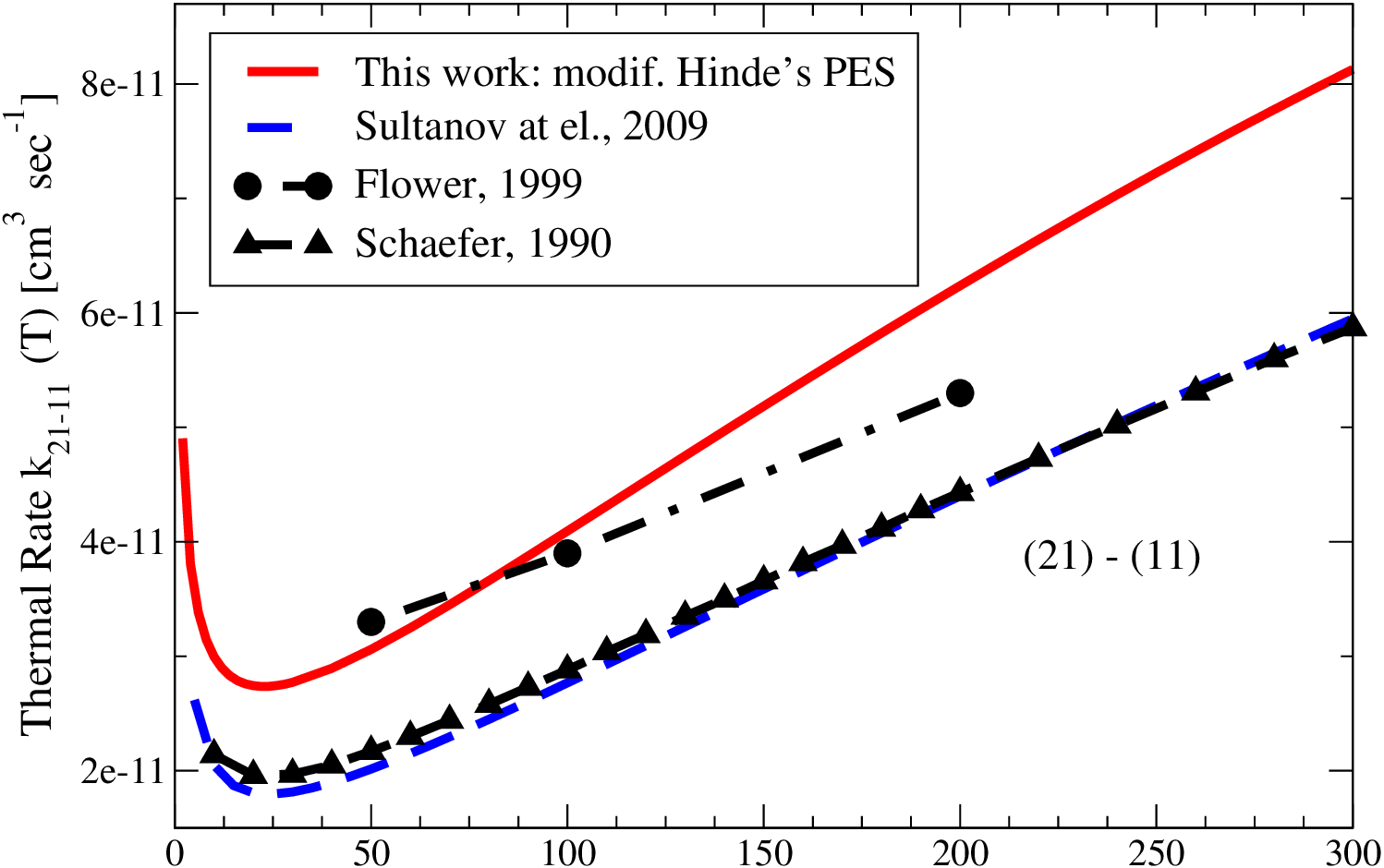}\vspace{4mm}\\
\includegraphics*[scale=1.0,width=20pc,height=13pc]{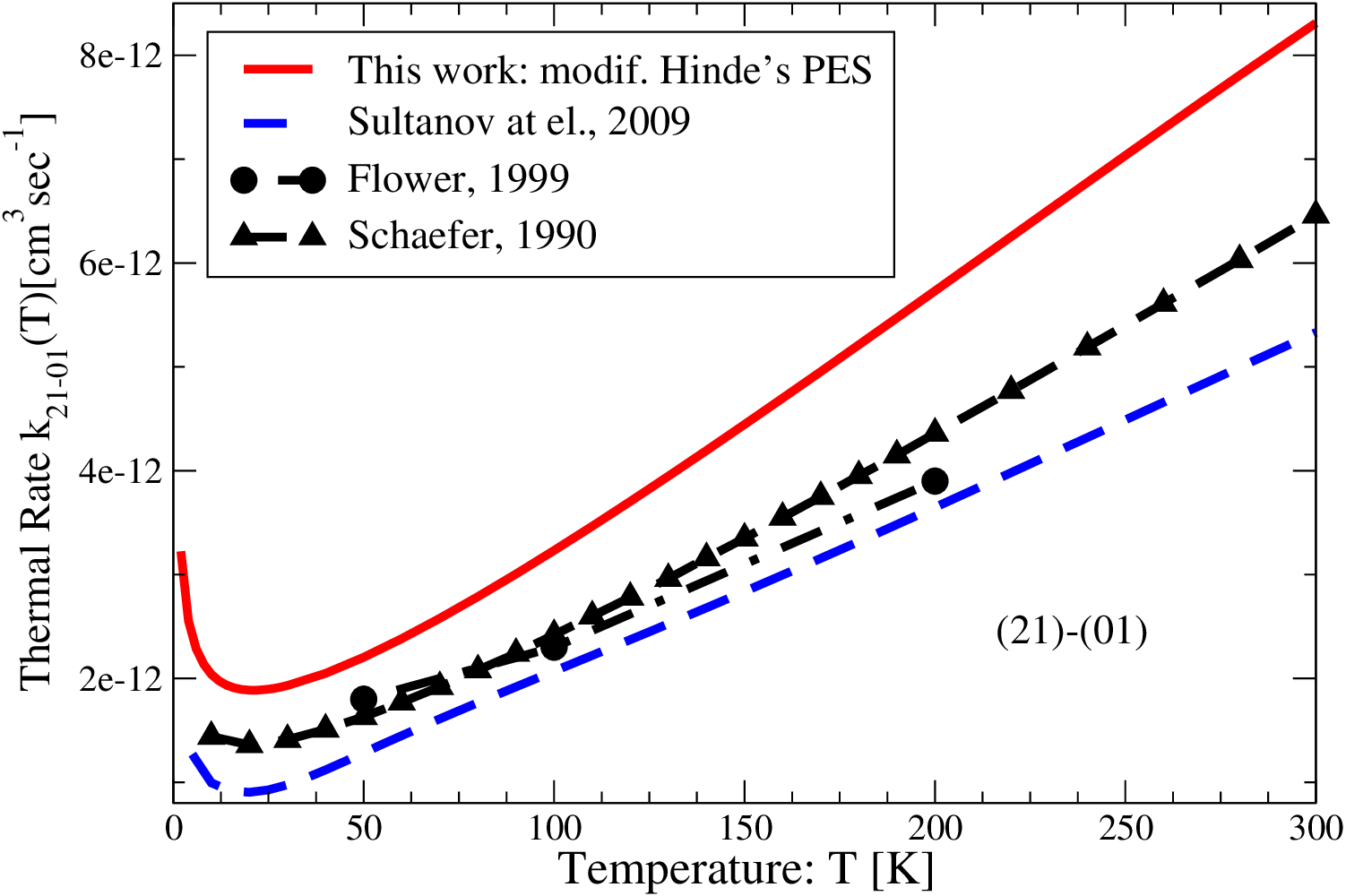}
\caption{(Color online) The HD+$ortho$-H$_2$ case.
Upper plot: the rotational de-excitation thermal rate coefficients $k_{ij\rightarrow i'j'}(T)$ for
the $\mbox{HD}(1) + \mbox{H}_2(1) \rightarrow \mbox{HD}(0) + \mbox{H}_2(1)$ collision.
The middle and lower plots represent $k_{ij\rightarrow i'j'}(T)$ for the
$\mbox{HD}(2) + \mbox{H}_2(1) \rightarrow \mbox{HD}(1) + \mbox{H}_2(1)$ and
$\mbox{HD}(2) + \mbox{H}_2(1) \rightarrow \mbox{HD}(0) + \mbox{H}_2(1)$ channels
correspondingly.}\label{fig:fig14}\end{center}\end{figure}

\begin{figure}\begin{center}            
\includegraphics*[scale=1.0,width=20pc,height=13pc]{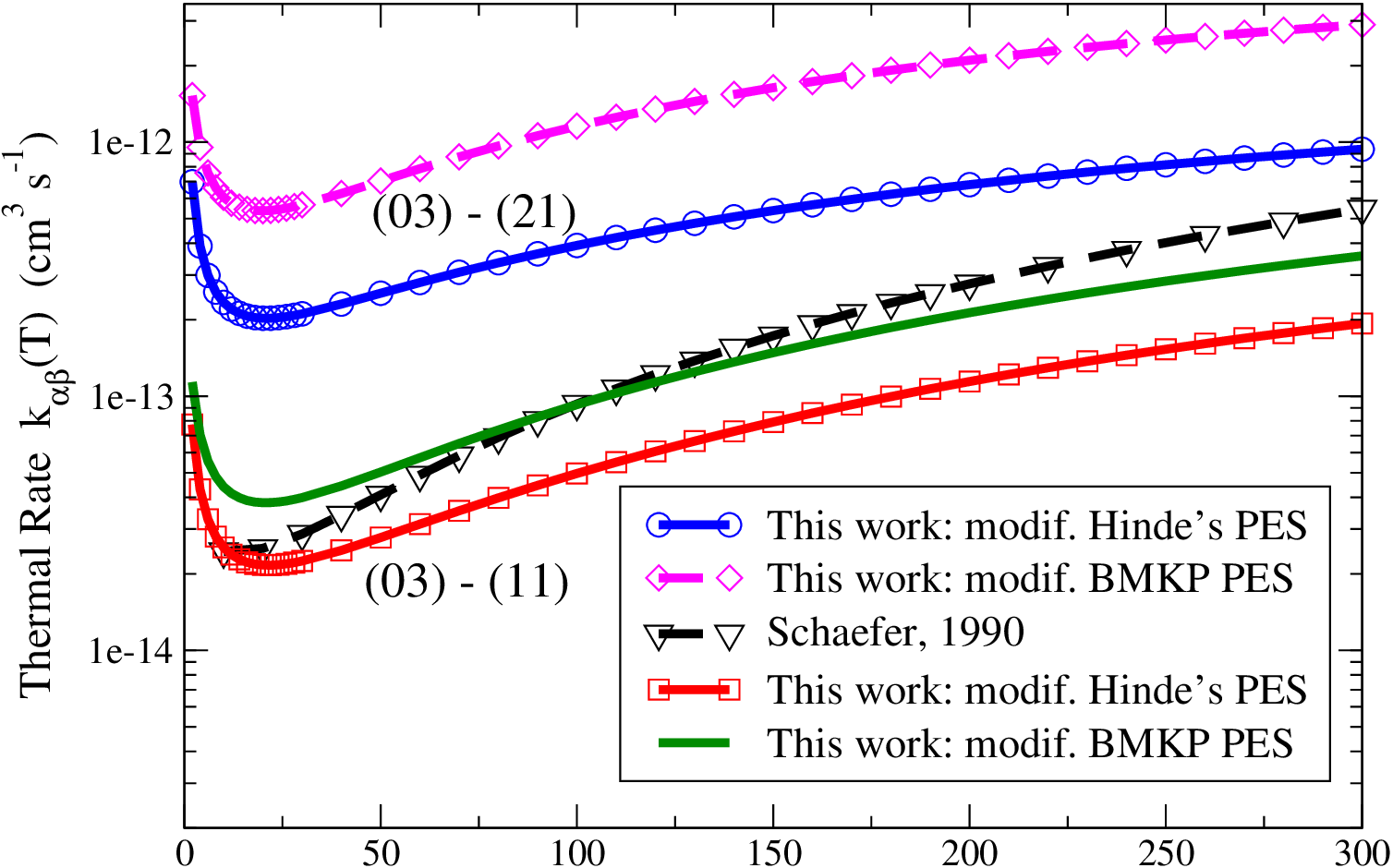}\vspace{8mm}\\
\includegraphics*[scale=1.0,width=20pc,height=13pc]{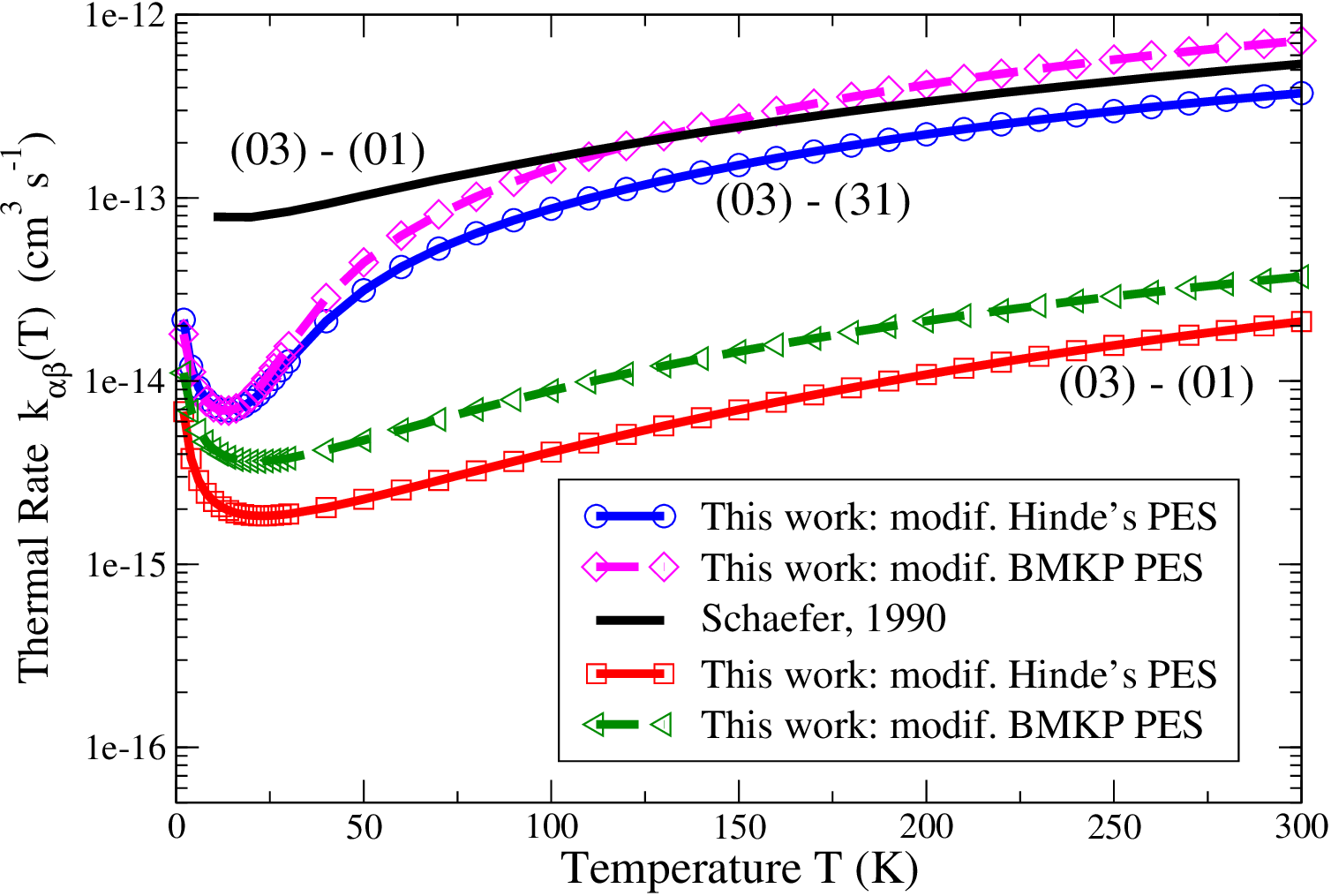} 
\caption{(Color online) The HD+$ortho$-H$_2$ case.
Upper plot: the rotational de-excitation thermal rate coefficients $k_{ij\rightarrow i'j'}(T)$ for the
$\mbox{HD}(0) + \mbox{H}_2(3) \rightarrow \mbox{HD}(2) + \mbox{H}_2(1)$ 
and 
$\mbox{HD}(0) + \mbox{H}_2(3) \rightarrow \mbox{HD}(1) + \mbox{H}_2(1)$ collisions
together with the corresponding result from \cite{schaefer}.
The lower plot represents $k_{ij\rightarrow i'j'}(T)$ for the
$\mbox{HD}(0) + \mbox{H}_2(3) \rightarrow \mbox{HD}(3) + \mbox{H}_2(1)$ and
$\mbox{HD}(0) + \mbox{H}_2(3) \rightarrow \mbox{HD}(0) + \mbox{H}_2(1)$ channels
correspondingly. The results of this paper were obtained with the use of the modified Hinde and BMKP PESs.}
\label{fig:fig15}\end{center}\end{figure}

Further, the following four Figs. (\ref{fig:fig8})-(\ref{fig:fig11}) represent our integral cross sections and corresponding
results from work\cite{schaefer} for the following de-excitation collisions: 
(\ref{eq:1202}), (\ref{eq:1220}), (\ref{eq:1210}), and (\ref{eq:1200}). In all of these processes
the initial state of HD has the rotational state $j_1$=1 and H$_2$ has the rotational state $j_2$=2. 
The corresponding total rotational energy of the molecules can be found in Table~\ref{tab:tab1}.
The results in Fig. (\ref{fig:fig8}) are in fairly good agreement with each other and
have relatively large values, therefore this specific channel could also make a substantial contribution to the
cooling function. It was found, that in Figs. (\ref{fig:fig9})-(\ref{fig:fig11}) both surfaces, namely, Hinde's
and BMKP both accurately reproduce the position of a shape resonance at collision velocity $v\sim$1300 m/sec.
Fig. \ref{fig:fig12} shows the resulting de-excitation cross sections from the highly located rotational quantum level. Finally, our
analysis in this paper would not be complete if we did not undertake computations for the $ortho$ hydrogen case as well. 
Fig. \ref{fig:fig13} shows results for the lower lying rotational quantum numbers of HD, 
namely, $j_1$=0 and 1, and H$_2$: $j_2$=1. It is seen that in this
case the cross sections obtained with the Hinde PES are significantly larger than other results. 
The corresponding thermal rate coefficients are presented in Fig. \ref{fig:fig14}. 
Again, as in previous $para$-hydrogen cases the new 
rates obtained with Hinde's potential are larger than previous results.
At the low density limit and taking into account the critical density concept the total cooling function can be computed
with the use of the following formula:
\begin{eqnarray}
\Lambda_{\rm{HD}} (T) = \sum_{j_1j_2,j'_1j'_2}
n_{\rm{HD}}(j_1)n_{\rm{H_2}}(j_2)k_{j_1j_2\rightarrow j'_1j'_2}(T) 
h\nu_{j_1j_2\rightarrow j'_1j'_2},
\label{eq:coolformula}
\end{eqnarray}
which is in units of [erg$\times$cm$^{-3}\times$s$^{-1}$].
Here, $h\nu_{j_1j_2\rightarrow j'_1j'_2}$ is the emitted photon energy,
$k_{j_1j_2\rightarrow j'_1j'_2}(T)$ it the thermal rate coefficient (\ref{eq:kT}) 
corresponding to the rotational transitions $j_1j_2\rightarrow j'_1j'_2$.
Therefore, increasing the knowledge of rotational and possibly vibrational
excitation and de-excitation rate constants, $k_{jv\rightarrow j'v'}(T)$,
in atomic and molecular hydrogen-hydrogen collisions, such as HD/H$_2$+H$_2$,
HD/H$_2$+H etc, is important in order to
understand and be able to model the energy balance in the interstellar medium.
For comparison purposes
it would be very useful and interesting to carry out new computations of the
rotational-vibrational integral cross sections and corresponding thermal rate coefficients for a low-energy HD+H collision.
In this case a different H$_3$ PESs from papers\cite{1peshd-h,2peshd-h} could be applied.

\acknowledgments
This paper was supported by the Office of Research and Sponsored Programs of
St. Cloud State University, USA and CNPq and FAPESP of Brazil.

\section{Appendix A: HD-H$_2$ potential energy surfaces}
\label{sec:HD_PES}
A few important modifications to the Hinde H$_2$-H$_2$ PES \cite{hinde}
were needed for the current non-symmetrical four-atomic collision (\ref{eq:h2hd}). 
The application and modification of the original H$_2$-H$_2$ BMKP PES were published in Ref. \cite{my2}.
Below in this paragraph we briefly describe the procedure.
To compute the distances between the four atoms 
the BMKP PES uses Cartesian coordinates. Consequently, it was necessary to convert spherical coordinates used in
the close-coupling method to the corresponding Cartesian coordinates and compute
the distances between the four atoms followed by calculation of the PES\cite{my2,my22}.
This procedure used a specifically oriented coordinate system $OXYZ$. 
As a first step we needed to introduce the Jacobi coordinates $\{\vec R_1, \vec R_2, \vec R_3\}$ and the radius-vectors 
of all four atoms in the space-fixed coordinate system 
$OXYZ$: $\{\vec r_1, \vec r_2, \vec r_3, \vec r_4\}$. 
Then the center of mass of the HD molecule has been relocated at the origin of the coordinate system $OXYZ$,
and the $\vec R_3$ was directed to center of mass of the H$_2$ molecule along the $OZ$ axis. 
Thus, one could obtain the following coordinate relationships:
$\vec R_3=\{R_3, \Theta_3=0, \Phi_3=0\}$, with $\Theta_3$ and $\Phi_3$ the polar and azimuthal angles,
$\vec R_1=\vec r_1 - \vec r_2$, $\vec R_2=\vec r_4 - \vec r_3$, $\vec r_1=\xi \vec R_1$ and $\vec r_2=(1 - \xi)
 \vec R_1$, where $\xi=m_2/(m_1+m_2)$ \cite{my2}.
Further, we adopted the $OXYZ$ system in such a way, that
the HD inter-atomic vector $\vec R_1$ lies on the $XOZ$ plane. Then the angle variables of
$\vec R_1$ and $\vec R_2$ are: $\hat R_1=\{\Theta_1,\Phi_1=\pi\}$
and $\hat R_2=\{\Theta_2,\Phi_2\}$ respectively.
One can see, that the Cartesian coordinates of the atoms of the HD molecule are \cite{my2}:
$\vec r_1 = \{x_1=\xi R_1\sin\Theta_1, y_1=0, z_1=\xi R_1 \cos\Theta_1\},\ \ \ 
\vec r_2 = \{x_2=-(1-\xi) R_1 \sin\Theta_1, y_2=0, 
z_2 = -(1-\xi) R_1 \cos\Theta_1\}.$
Defining  $\zeta = m_4/(m_3+m_4)$, we have
$
\vec r_3 = \vec R_3 - (1-\zeta)\vec R_2,\ \ \  \vec r_4 = \vec R_3 + \zeta \vec R_2,
$
and the corresponding Cartesian coordinates are:
$
\vec r_3 = \{x_3 = - (1-\zeta) R_2 \sin\Theta_2 \cos\Phi_2,\ \ \      
y_3 = - (1-\zeta)R_2\sin\Theta_2\sin\Phi_2,\ \ \                             
z_3 = R_3-(1-\zeta)R_2\cos\Theta_2\},
$
$
\vec r_4 = \{x_4 = \zeta R_2 \sin\Theta_2 \cos\Phi_2,\ \ \    
y_4 = \zeta R_2 \sin\Theta_2\sin\Phi_2,\ \ \                         
z_4 = R_3+\zeta R_2\cos\Theta_2\}.
$
In such a manner the cartesian and the Jacobi coordinates are
represented together for the four-atomic system HD+H$_2$ \cite{my2}.

The Hinde H$_2$-H$_2$ PES\cite{hinde} is a
six-dimensional surface which was constructed using  recent Raman spectrum data of the
(H$_2$)$_2$ dimer and  which accurately describes the dimer's van der
Waals well \cite{hinde}.
It was demonstrated that this PES  gives IR and Raman transition energies
for the ($para$-H$_2$)$_2$, ($ortho$-D$_2$)$_2$, and ($para$-H$_2$)$-$($ortho$-D$_2$)
dimers and is in good agreement with experimental data.
 
The method to make the Hinde H$_2$-H$_2$ PES 
suitable for the non-symmetric system HD+H$_2$
is based on a geometrical
transformation technique, i.e. a  rotation of the three-dimensional space and the corresponding space-fixed coordinate
system $OXYZ$. The new global PES depends on six variables (Jacobi coordinates) $-$ $|\vec R_1|$, $|\vec R_2|$, 
$|\vec R_3|$, $\theta_1$, $\theta_2$, and $\varphi_2$ $-$  as shown in Fig.~\ref{fig:fig1}
together with corresponding quantum angular momenta.
The initial geometry of the system is designed in such a way that 
the Jacobi vector $\vec R_3$ connects the c.m.'s of the two H$_2$ molecules and is
directed over the ${OZ}$ axis. We laid out $OXYZ$ in such a manner 
that the Jacobi vector $\vec R_1$ lies in the $XOZ$ plane.
The vector $\vec R_2$ can then be directed anywhere. The spherical coordinates of the Jacobi vectors are:
$\vec R_1 = \{R_1, \theta_1, 0\}$, $\vec R_2 = \{R_2, \theta_2, \varphi_2\}$,  and $\vec R_3 = \{R_3, 0, 0\}$.
Because we used the rigid rotor
model, the lengths of the HD and H$_2$ molecules are fixed at equilibrium values, e.g., 
$R_1$=0.7631 a.u. and $R_2$=0.7668 a.u., thus leaving us with  four free variables.
We replace one hydrogen atom H with a deuterium atom D, thus
shifting the c.m. of one H$_2$ molecule 
to another point, that is from $O$ to $O^{\prime}$. 
The length of the vector $\vec x$  is $x = |\vec R_{1}|/6$.
This is seen in Fig.~\ref{fig:fig223}. 
Then we shift the original coordinate system $OXYZ$ 
along the vector $\vec R_1$ to 
the new one $O'X'Y'Z'$. The origin of the new system, i.e. $O'$, 
lies on the c.m. of HD. 
A new Jacobi vector  $\vec R'_3$ is defined
 connecting the c.m.'s of the HD and H$_2$ molecules.
The new intermolecular
distance between HD and H$_2$ is:
\begin{eqnarray}\label{eq:r3prime}
R'_3 = \sqrt{R_3^2 + x^2 - 2 x R_3 \cos \theta_1}.
\end{eqnarray}
Now, if we rotate $O'X'Y'Z'$ around its ${O'Y'}$ axis in such a way that the ${OZ'}$ 
axis is directed over the vector $\vec R'_3$ we obtain a new coordinate system $O'X''Y''Z''$
which should be well designed to carry out computations
for the HD+H$_2$ collision.  The rotational angle $\eta$ satisfies:
\begin{eqnarray}\label{eq:etta1}
\cos \eta &=& \frac{R'^2_3 + R_3^2 - x^2}{2 R'_3 R_3}\\
\sin \eta  &=& \frac{x}{R'_3 \sin \theta_1}
\label{eq:etta2}\end{eqnarray}
This transformation converts the initial Jacobi vectors in $OXYZ$:
$\vec R_1 = \{R_1, \theta_1, 0\}$, $\vec R_2 = \{R_2, \theta_2, \varphi_2\}$ and $\vec R_3 = \{R_3, 0, 0\}$ 
to the corresponding Jacobi vectors with new coordinates in the
new $O'X'Y'Z'$: $\vec R'_1 = \{R'_1, \theta'_1, \varphi'_{2}\}$,
$\vec R'_2 = \{R'_2, \theta'_2, 0\}$ and $\vec R'_3 = \{R'_3, 0, 0\}$.

The coordinate transformations from $OXYZ$ to $O''X''Y'´Z''$ changes the original Hinde's PES
to the new HD-H$_2$ PES. Rotation of the 
coordinate system from $O'X'Y'Z'$ to $O'X''Y''Z''$ results in a corresponding transformation of the
coordinates of the 4-body system as well as changing the distance between the two molecules.
One then has the following relations between new and old variables \cite{varshal88}:
\begin{eqnarray}\label{eq:RX}
\theta'_{1}   &=& \arccos(\cos \theta_1 \cos \eta + \sin \theta_1 \sin \eta),\\  
\theta'_{2}   &=& \arccos(\cos \theta_2 \cos \eta + \sin \theta_2 \sin \eta \cos \Phi_2),\\
\varphi'_{2} &=& \arccos \left ( \cot \phi_2 \cos \eta + \frac{\cot \theta_1 \sin \eta}  {\sin \phi_{2}} \right ).
\label{eq:R3}
\end{eqnarray}
In the calculation of HD+H$_2$ with Hinde's PES one has to use new 
coordinates $\theta'_{1}, \theta'_{2}, \varphi'_{2}, R'_{3}$. However, the original
potential has been expressed through the initial H$_2$-H$_2$
variables, i.e. $\theta_1$, $\theta_2$, $\Phi_2$ and $R_3$. Hence they have
to be transformed using (\ref{eq:RX})-(\ref{eq:R3}). Therefore, in the case of the
non-symmetrical HD+H$_2$ collision one should use the formulas (\ref{eq:RX})-(\ref{eq:R3})
together with (\ref{eq:etta1})-(\ref{eq:etta2}) and the expression (\ref{eq:r3prime}) for the new 
distance $R^{\prime}_3$ between the center of masses of the H$_2$ and HD molecules.

In general, any consideration of the HD+H$_2$, D$_2$+HD or D$_2$+D$_2$
systems should begin with the original H$_2$-H$_2$ PES.
This six-dimensional function comprises a symmetrical surface over
the $OZ$ coordinate axis. This is shown in Figs.\ 1 and 16. In spherical coordinates 
the surface can be described by
six variables: $R_1$, $\theta_1$, $R_2$, $\theta_2$, $R_3$, and $\varphi_2$.
The variables are also shown in the figures. The H$_2$-H$_2$ PES was obtained in
the framework of the Born-Oppenheimer model \cite{BO} and can be
considered as a symmetrical interaction field. When considering non-reactive
scattering problems with participation of hydrogen
molecules one needs to solve the Schr\"odinger Eq. (\ref{eq:schred})
with the H$_2$-H$_2$ potential $V(\vec R_1,\vec R_2,\vec R_3)$.
The solution/propagation runs over the $\vec R_3$ Jacobi vector 
(please see Fig.\ 16). Therefore,
in the case of the symmetrical H$_2$+H$_2$ and D$_2$+D$_2$
collisions one can use the original H$_2$-H$_2$ PES as it is,
i.e. without transformations.

However, in the case of the non-symmetrical (or symmetry-broken) HD+H2/D2 or HD+HD scattering
systems one should also apply the original H$_2$-H$_2$ interaction field (PES),
but the propagation (solution) of the Schr\"odinger equation runs, in this case,
over the corrected Jacobi vector $\vec R'_3$ which is directed over the new
$OZ''$ axis, as is shown in Fig.\ 16.  

\section{Appendix B: HD+$o$-/$p$-H$_2$ thermal rate coefficients}
\label{sec:rates7}
New data for the thermal rate coefficients, $k_{j_1j_2\rightarrow j'_1j'_2}(T)$, Eq. (\ref{eq:kT}),
are listed below. The results are obtained with the modified Hinde H$_2$-H$_2$ PES \cite{hinde}:
Table~\ref{tab:tab2} includes $k_{j_1j_2\rightarrow j'_1j'_2}(T)$ at low temperatures
for the HD + $para$-H$_2$ rotational energy transfer collisions and
Table~\ref{tab:tab3} the same information for the HD + $ortho$-H$_2$ case.
The astrophysical HD-cooling function can be 
computed with the use of formula
(\ref{eq:coolformula}).  

\clearpage
\begin{table}        
\caption{Total rotational energy $\mathcal{E}_{rot}$ in the four-atomic system: HD($j_1$)+$p$-H$_2(j_2)$
and HD($j_1$)+$o$-H$_2(j_2)$. Here, $\mathcal{E}_{rot} = B_1j_1(j_1+1) + B_2j_2(j_2+1)$, 
where $B_{1(2)}$ are the rotational constants of rigid rotors $ab$ and $cd$ respectively.}
\label{tab:tab1}
\vspace{2mm}
\centering\begin{ruledtabular}\begin{tabular}{lcclcc}
$\mathcal{E}_{rot}$ (cm$^{-1}$)&HD($j_1$)&$p$-H$_2(j_2)$  &  $\mathcal{E}_{rot}$ (cm$^{-1}$)&HD($j_1$)&$o$-H$_2(j_2)$\\
\hline
0.0      &0&0&   121.6    &0&1\\ 
89.4    &1&0&   211.0    &1&1\\ 
268.2  &2&0&   389.8    &2&1\\ 
364.8  &0&2&   658.0    &3&1\\ 
454.2  &1&2&   729.6    &0&3\\
536.4  &3&0&   819.0    &1&3\\ 
633.0  &2&2&   997.8    &2&3
\end{tabular}\end{ruledtabular}\end{table}

\clearpage
\begin{table*}            
\caption{Low temperature rotational de-excitation thermal rate coefficients $k_{ij\rightarrow i'j'}(T)$
in the $\mbox{HD}(i) + para$-$\mbox{H}_2(j) \rightarrow \mbox{HD}(i') + para$-$\mbox{H}_2(j')$ collision.
All results are multiplied by a constant value $\alpha=10^{11}$. The data are in the unit cm$^3$ sec$^{-1}$.}
\label{tab:tab2}
\vspace{2mm}
\centering
\begin{ruledtabular}
\begin{tabular}{lcccccccccc}
$T$(K) & 10-00 & 20-10 & 20-00 & 02-20 & 02-10 & 02-00 & 12-02 & 12-20 & 12-10 & 12-00\\
\hline
    2  & 4.774  & 6.892  & 1.115  & 0.914  &3.88E-02  &2.45E-03  & 4.960  &2.22E-02  &5.61E-03  &6.01E-04\\
    4  & 4.099  & 4.817  & 0.670  & 0.553  &2.31E-02  &1.46E-03  & 4.194  &1.74E-02  &4.40E-03  &4.56E-04\\
    6  & 3.780  & 4.007  & 0.485  & 0.444  &1.83E-02  &1.16E-03  & 3.841  &1.55E-02  &3.96E-03  &4.04E-04\\
    8  & 3.574  & 3.580  & 0.390  & 0.393  &1.60E-02  &1.01E-03  & 3.617  &1.44E-02  &3.74E-03  &3.76E-04\\
   10  & 3.431  & 3.321  & 0.335  & 0.363  &1.47E-02  &9.26E-04  & 3.464  &1.38E-02  &3.61E-03  &3.60E-04\\
   12  & 3.329  & 3.153  & 0.299  & 0.345  &1.39E-02  &8.74E-04  & 3.356  &1.33E-02  &3.54E-03  &3.51E-04\\
   14  & 3.256  & 3.039  & 0.274  & 0.333  &1.34E-02  &8.39E-04  & 3.278  &1.31E-02  &3.51E-03  &3.45E-04\\
   16  & 3.203  & 2.960  & 0.257  & 0.325  &1.31E-02  &8.17E-04  & 3.223  &1.29E-02  &3.50E-03  &3.43E-04\\
   18  & 3.166  & 2.907  & 0.245  & 0.319  &1.29E-02  &8.03E-04  & 3.183  &1.29E-02  &3.50E-03  &3.41E-04\\
   20  & 3.140  & 2.872  & 0.236  & 0.315  &1.27E-02  &7.94E-04  & 3.157  &1.32E-02  &3.52E-03  &3.41E-04\\
   22  & 3.125  & 2.850  & 0.229  & 0.313  &1.27E-02  &7.91E-04  & 3.140  &1.41E-02  &3.54E-03  &3.43E-04\\
   24  & 3.116  & 2.839  & 0.224  & 0.311  &1.27E-02  &7.91E-04  & 3.131  &1.63E-02  &3.57E-03  &3.47E-04\\
   26  & 3.114  & 2.836  & 0.221  & 0.310  &1.28E-02  &7.94E-04  & 3.129  &2.04E-02  &3.62E-03  &3.53E-04\\
   30  & 3.124  & 2.848  & 0.217  & 0.310  &1.30E-02  &8.06E-04  & 3.141  &3.84E-02  &3.82E-03  &3.82E-04\\
   40  & 3.204  & 2.948  & 0.218  & 0.313  &1.39E-02  &8.66E-04  & 3.228  &1.78E-01  &5.29E-03  &6.11E-04\\
   50  & 3.327  & 3.102  & 0.227  & 0.319  &1.51E-02  &9.48E-04  & 3.360  &4.83E-01  &8.52E-03  &1.12E-03\\
   60  & 3.474  & 3.283  & 0.241  & 0.325  &1.64E-02  &1.05E-03  & 3.512  &9.10E-01  &1.32E-02  &1.84E-03\\
   80  & 3.798  & 3.687  & 0.275  & 0.335  &1.93E-02  &1.27E-03  & 3.841  &1.86E+00  &2.39E-02  &3.49E-03\\
  100  & 4.137  & 4.115  & 0.315  & 0.342  &2.24E-02  &1.52E-03  & 4.178  &2.66E+00  &3.34E-02  &4.95E-03\\
  120  & 4.478  & 4.549  & 0.359  & 0.347  &2.54E-02  &1.81E-03  & 4.514  &3.21E+00  &4.07E-02  &6.05E-03\\
  140  & 4.815  & 4.982  & 0.405  & 0.351  &2.85E-02  &2.11E-03  & 4.844  &3.54E+00  &4.59E-02  &6.82E-03\\
  160  & 5.145  & 5.410  & 0.452  & 0.353  &3.16E-02  &2.45E-03  & 5.168  &3.72E+00  &4.95E-02  &7.35E-03\\
  180  & 5.469  & 5.830  & 0.500  & 0.354  &3.47E-02  &2.80E-03  & 5.485  &3.78E+00  &5.20E-02  &7.72E-03\\
  200  & 5.783  & 6.241  & 0.550  & 0.354  &3.77E-02  &3.17E-03  & 5.793  &3.77E+00  &5.38E-02  &7.99E-03\\
  250  & 6.528  & 7.218  & 0.674  & 0.352  &4.50E-02  &4.18E-03  & 6.525  &3.58E+00  &5.65E-02  &8.45E-03\\
  300  & 7.204  & 8.114  & 0.796  & 0.349  &5.18E-02  &5.23E-03  & 7.193  &3.29E+00  &5.81E-02  &8.81E-03\\
\end{tabular}
\end{ruledtabular}
\end{table*}

\clearpage
\begin{table*}            
\caption{Low temperature rotational de-excitation thermal rate coefficients $k_{ij\rightarrow i'j'}(T)$
in the $\mbox{HD}(i) + ortho$-$\mbox{H}_2(j) \rightarrow \mbox{HD}(i') + ortho$-$\mbox{H}_2(j')$ collision.
All results are multiplied by a constant value $\alpha=10^{11}$. The data are in the unit cm$^3$ sec$^{-1}$.}
\label{tab:tab3}
\vspace{2mm}
\centering
\begin{ruledtabular}
\begin{tabular}{lccccccccccc}
$T$(K) & 11-01       & 21-11      & 21-01      & 31-21      & 31-11      & 31-01      & 03-21      & 03-11      & 03-31      & 13-03      & 13-31  \\
\hline
    2  & 4.921  & 4.903  & 0.322  & 3.691  & 0.299  &3.68E-02  &6.98E-02  &7.74E-03  &2.16E-03  & 5.037  &1.50E-01\\
    4  & 4.189  & 3.807  & 0.256  & 2.669  & 0.218  &2.69E-02  &3.90E-02  &4.30E-03  &1.20E-03  & 4.222  &1.10E-01\\
    6  & 3.844  & 3.383  & 0.229  & 2.304  & 0.189  &2.33E-02  &2.99E-02  &3.28E-03  &9.19E-04  & 3.855  &9.58E-02\\
    8  & 3.623  & 3.146  & 0.214  & 2.112  & 0.173  &2.14E-02  &2.57E-02  &2.80E-03  &7.94E-04  & 3.625  &8.88E-02\\
   10  & 3.472  & 2.997  & 0.204  & 1.996  & 0.164  &2.03E-02  &2.34E-02  &2.54E-03  &7.31E-04  & 3.469  &8.48E-02\\
   12  & 3.364  & 2.898  & 0.198  & 1.922  & 0.158  &1.96E-02  &2.20E-02  &2.38E-03  &7.00E-04  & 3.359  &8.24E-02\\
   14  & 3.287  & 2.831  & 0.194  & 1.874  & 0.154  &1.91E-02  &2.12E-02  &2.28E-03  &6.92E-04  & 3.280  &8.10E-02\\
   16  & 3.231  & 2.787  & 0.191  & 1.843  & 0.152  &1.88E-02  &2.07E-02  &2.22E-03  &7.02E-04  & 3.223  &8.03E-02\\
   18  & 3.192  & 2.759  & 0.189  & 1.825  & 0.151  &1.87E-02  &2.04E-02  &2.18E-03  &7.30E-04  & 3.183  &8.00E-02\\
   20  & 3.164  & 2.743  & 0.189  & 1.816  & 0.150  &1.87E-02  &2.03E-02  &2.17E-03  &7.77E-04  & 3.155  &8.00E-02\\
   22  & 3.147  & 2.737  & 0.189  & 1.814  & 0.150  &1.87E-02  &2.03E-02  &2.16E-03  &8.42E-04  & 3.137  &8.04E-02\\
   24  & 3.138  & 2.737  & 0.189  & 1.817  & 0.151  &1.88E-02  &2.04E-02  &2.17E-03  &9.26E-04  & 3.127  &8.09E-02\\
   26  & 3.134  & 2.744  & 0.190  & 1.826  & 0.152  &1.89E-02  &2.06E-02  &2.19E-03  &1.03E-03  & 3.124  &8.16E-02\\
   30  & 3.143  & 2.772  & 0.193  & 1.853  & 0.155  &1.94E-02  &2.11E-02  &2.25E-03  &1.28E-03  & 3.131  &8.33E-02\\
   40  & 3.220  & 2.895  & 0.205  & 1.962  & 0.166  &2.10E-02  &2.30E-02  &2.47E-03  &2.12E-03  & 3.207  &8.87E-02\\
   50  & 3.343  & 3.060  & 0.220  & 2.105  & 0.181  &2.31E-02  &2.54E-02  &2.78E-03  &3.13E-03  & 3.328  &9.50E-02\\
   60  & 3.488  & 3.249  & 0.238  & 2.269  & 0.199  &2.56E-02  &2.80E-02  &3.14E-03  &4.19E-03  & 3.471  &1.02E-01\\
   80  & 3.811  & 3.661  & 0.279  & 2.634  & 0.240  &3.16E-02  &3.35E-02  &3.98E-03  &6.40E-03  & 3.792  &1.14E-01\\
  100  & 4.150  & 4.094  & 0.323  & 3.028  & 0.287  &3.88E-02  &3.92E-02  &4.96E-03  &8.72E-03  & 4.127  &1.26E-01\\
  120  & 4.490  & 4.532  & 0.370  & 3.440  & 0.339  &4.69E-02  &4.50E-02  &6.07E-03  &1.12E-02  & 4.465  &1.37E-01\\
  140  & 4.827  & 4.969  & 0.419  & 3.864  & 0.396  &5.60E-02  &5.08E-02  &7.28E-03  &1.38E-02  & 4.799  &1.46E-01\\
  160  & 5.157  & 5.400  & 0.470  & 4.295  & 0.457  &6.60E-02  &5.66E-02  &8.58E-03  &1.65E-02  & 5.127  &1.55E-01\\
  180  & 5.480  & 5.824  & 0.521  & 4.731  & 0.521  &7.69E-02  &6.24E-02  &9.97E-03  &1.93E-02  & 5.447  &1.62E-01\\
  200  & 5.795  & 6.238  & 0.573  & 5.169  & 0.589  &8.87E-02  &6.80E-02  &1.14E-02  &2.22E-02  & 5.759  &1.69E-01\\
  250  & 6.539  & 7.224  & 0.704  & 6.258  & 0.772  &1.21E-01  &8.14E-02  &1.53E-02  &2.98E-02  & 6.498  &1.82E-01\\
  300  & 7.215  & 8.127  & 0.831  & 7.311  & 0.964  &1.57E-01  &9.37E-02  &1.93E-02  &3.73E-02  & 7.171  &1.92E-01\\
\end{tabular}
\end{ruledtabular}
\end{table*}


\clearpage
\clearpage

\clearpage
\clearpage
\clearpage
\clearpage
\clearpage

\clearpage
\clearpage
\clearpage
\clearpage
\clearpage
\clearpage
\clearpage

\clearpage

\clearpage

\begin{figure}\begin{center}            
\includegraphics*[scale=1.0,width=20pc,height=13pc]{13_0331211101RHkT.eps}
\caption{(Color online) The HD+$ortho$-H$_2$ case.
Upper plot: the rotational de-excitation thermal rate coefficients $k_{ij\rightarrow i'j'}(T)$ for
the $\mbox{HD}(1) + \mbox{H}_2(3) \rightarrow \mbox{HD}(0) + \mbox{H}_2(3)$ collision.
The middle plot represents $k_{ij\rightarrow i'j'}(T)$ for the
$\mbox{HD}(1) + \mbox{H}_2(3) \rightarrow \mbox{HD}(3) + \mbox{H}_2(1)$ process,
and lower plot shows three results for the
$\mbox{HD}(1) + \mbox{H}_2(3) \rightarrow \mbox{HD}(2) + \mbox{H}_2(1)$,
$\mbox{HD}(1) + \mbox{H}_2(3) \rightarrow \mbox{HD}(1) + \mbox{H}_2(1)$, and
$\mbox{HD}(1) + \mbox{H}_2(3) \rightarrow \mbox{HD}(0) + \mbox{H}_2(1)$ collisions
correspondingly. The results of this paper were obtained with the use of the modified Hinde and BMKP PESs.}
\label{fig:fig16}\end{center}\end{figure}
\clearpage

\begin{figure}\begin{center}            
\includegraphics[scale=.42]{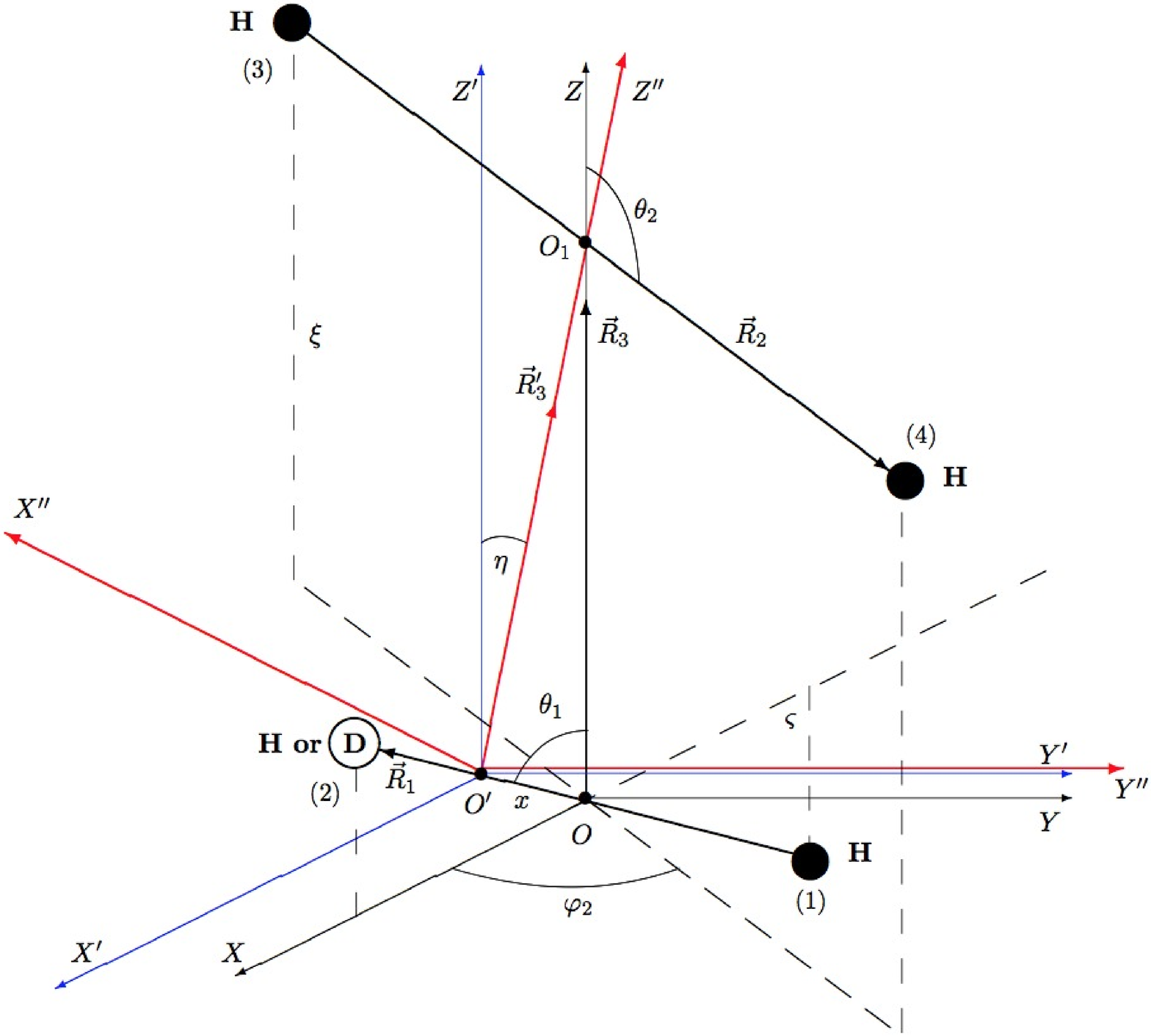}
\caption{(Color online) Four-atomic system 1234 or H-D-H-H is shown together with its few-body Jacobi
coordinates $\{\vec R_1, \vec R_2, \vec R_3\}$. The original cartesian coordinate system is $OXYZ$.
The center of mass of the original H$_2$ molecule lies in point $O$.
$O'X'Y'Z'$ is a system which was shifted in parallel from the original system, $O'$ lies in the center of mass
of the actual HD molecule. The close-coupling equations
are solved using the space-fixed coordinate system
$O'X''Y''Z''$. 
The vector $\vec R'_3$ connects the center of masses of
the HD and H$_2$ molecules.
HD is the first molecule with a rotational constant B$_e$(1)=44.7 cm$^{-1}$ and quantum angular momentum of $j_1$,
H$_2$ is the second molecule with rotational constant of B$_e$(2)=60.8 cm$^{-1}$ 
and uses momentum $j_2$ in the system;
$R_{1}=0.7631$ \r{A} and $R_{2}=0.7668$ \r{A}
are fixed interatomic distances in each hydrogen
molecule HD and H$_2$ respectively.}
\label{fig:fig223}
\end{center}
\end{figure}
\end{document}